\documentclass[twocolumn]{aastex631}

\pdfoutput=1 

\usepackage[utf8]{inputenc}
\usepackage{makecell}
\usepackage{amsmath}
\usepackage{tikz}
\usepackage{hyperref}
\usepackage{soul}
\usepackage{graphicx}
\usepackage{natbib}

\setstcolor{red}

\newcommand{\Oiii}{[{\sc Oiii}]}
\newcommand{\Muv}{M_\mathrm{UV}}
\newcommand{\vcir}{v_\mathrm{cir}}
\newcommand{\nel}{n_\mathrm{e}}
\newcommand{\vout}{v_\mathrm{out}}
\newcommand{\rout}{r_\mathrm{out}}

\defcitealias{Carniani+23}{C23}
\newcommand{\Car}{\citetalias{Carniani+23}}

\newcommand{\noedit}[1]{#1}

\received{---}
\revised{---}
\accepted{---}

\shorttitle{
Outflows in JWST Galaxies at $z=3-9$
}
\shortauthors{Xu et al.}

\begin{document}
\title{
Stellar- and AGN-Driven Outflows in JWST Galaxies at $z=3-9$:\\
More Frequent, Wider Opening Angles, and Mostly Bounded
}

\correspondingauthor{Yi Xu}
\email{xuyi@icrr.u-tokyo.ac.jp}

\author[0000-0002-5768-8235]{Yi Xu}
\affiliation{Institute for Cosmic Ray Research, The University of Tokyo, 5-1-5 Kashiwanoha, Kashiwa, Chiba 277-8582, Japan}
\affiliation{Department of Astronomy, Graduate School of Science, the University of Tokyo, 7-3-1 Hongo, Bunkyo, Tokyo 113-0033, Japan}

\author[0000-0002-1049-6658]{Masami Ouchi}
\affiliation{National Astronomical Observatory of Japan, 2-21-1 Osawa, Mitaka, Tokyo 181-8588, Japan}
\affiliation{Institute for Cosmic Ray Research, The University of Tokyo, 5-1-5 Kashiwanoha, Kashiwa, Chiba 277-8582, Japan}
\affiliation{Graduate University for Advanced Studies (SOKENDAI), 2-21-1 Osawa, Mitaka, Tokyo 181-8588, Japan}
\affiliation{Kavli Institute for the Physics and Mathematics of the Universe (Kavli IPMU, WPI), The University of Tokyo, 5-1-5 Kashiwanoha, Kashiwa, Chiba, 277-8583, Japan}

\author[0000-0003-2965-5070]{Kimihiko Nakajima}
\affiliation{National Astronomical Observatory of Japan, 2-21-1 Osawa, Mitaka, Tokyo 181-8588, Japan}

\author[0000-0002-6047-430X]{Yuichi Harikane}
\affiliation{Institute for Cosmic Ray Research, The University of Tokyo, 5-1-5 Kashiwanoha, Kashiwa, Chiba 277-8582, Japan}

\author[0000-0001-7730-8634]{Yuki Isobe}
\affiliation{Institute for Cosmic Ray Research, The University of Tokyo, 5-1-5 Kashiwanoha, Kashiwa, Chiba 277-8582, Japan}
\affiliation{Department of Physics, Graduate School of Science, The University of Tokyo, 7-3-1 Hongo, Bunkyo, Tokyo 113-0033, Japan}

\author[0000-0001-9011-7605]{Yoshiaki Ono}
\affiliation{Institute for Cosmic Ray Research, The University of Tokyo, 5-1-5 Kashiwanoha, Kashiwa, Chiba 277-8582, Japan}

\author[0009-0008-0167-5129]{Hiroya Umeda}
\affiliation{Institute for Cosmic Ray Research, The University of Tokyo, 5-1-5 Kashiwanoha, Kashiwa, Chiba 277-8582, Japan}
\affiliation{Department of Physics, Graduate School of Science, The University of Tokyo, 7-3-1 Hongo, Bunkyo, Tokyo 113-0033, Japan}

\author[0000-0003-3817-8739]{Yechi Zhang}
\affiliation{Institute for Cosmic Ray Research, The University of Tokyo, 5-1-5 Kashiwanoha, Kashiwa, Chiba 277-8582, Japan}
\affiliation{Department of Astronomy, Graduate School of Science, the University of Tokyo, 7-3-1 Hongo, Bunkyo, Tokyo 113-0033, Japan}
\affiliation{Kavli Institute for the Physics and Mathematics of the Universe (Kavli IPMU, WPI), The University of Tokyo, 5-1-5 Kashiwanoha, Kashiwa, Chiba, 277-8583, Japan} 
\affiliation{IPAC, California Institute of Technology, MC 314-6, 1200 E. California Boulevard, Pasadena, CA 91125, USA}

\begin{abstract}
We study outflows in 130 galaxies with $-22<M_{\rm UV}<-16$ at $z=3-9$ identified in JWST NIRSpec and NIRCam WFSS data taken by the ERO, CEERS, FRESCO, GLASS, and JADES programs. We identify 30 out of the 130 galaxies with broad components of FWHM$\sim200-700~\mathrm{km~s^{-1}}$ in the emission lines of H$\alpha$ and \Oiii~that trace ionized outflows. Four out of the 30 outflowing galaxies are Type-1 AGN whose H$\alpha$ emission lines include line profile components as broad as FWHM$\gtrsim 1000~\mathrm{km~s^{-1}}$ while one galaxy is identified as a Type-2 AGN by high ionization emission lines. With the velocity shift and line widths of the outflow broad lines, we obtain $\sim 80-500~\mathrm{km~s^{-1}}$ for the outflow velocities. We find that the outflow velocities are slower than low-z galaxies with similar SFRs, which may be explained by the low stellar masses of high-z galaxies.
The outflow velocities of AGNs are large but not significantly different from the others. Interestingly, these outflow velocities are typically not high enough to escape from the galactic potentials, possibly suggesting fountain-type outflows. We estimate mass loading factors $\eta$ to be $0.1-1$ that are not particularly large, but comparable with those of $z\sim 1$ outflows. The large fraction of galaxies with outflows (30\% with high resolution data) provides constraints on outflow parameters, suggesting a wide opening angle of $\gtrsim 45$ deg and a large duty-cycle of $\gtrsim30$\%, which gives a picture of more frequent and spherical outflows in high-$z$ galaxies.

\end{abstract}

\keywords{galaxies: evolution --- galaxies: kinematics and dynamics}

\section{Introduction}
% 1. background on outflows
Current Cold Dark Matter (CDM) models predict well for the observed large-scale structure. However, the details of galaxy evolution cannot be well modelled without introducing the complicated baryonic processes one of which is the feedback from star formation or active galactic nuclei (AGN) activity. Recent observations with the James Webb Space Telescope (JWST) have identified several UV-luminous galaxies at $z\sim10$ \citep[][]{Naidu+22,Adams+23,Harikane+23a}, which suggests stellar mass buildup is efficient in the early universe. The existence of these UV-luminous galaxies can be explained by a top-heavy IMF \cite[e.g.,][]{Inayoshi+22}, the lack of suppression of star formation \citep[e.g.,][]{Dekel+23}, AGN activities, or high star formation efficiency \cite[e.g.,][]{Harikane+23c}. Among these explanations, mechanical and thermal energy injected by supernovae, massive stars, or AGNs are responsible for removing gas from the star formation region and suppressing star formation \citep{Veilleux+05}, which happens ubiquitously in the local universe \citep[][]{Rupke+05,Weiner+09} and possibly becomes stronger at high redshift \citep[][]{Sugahara+17,Sugahara+19}. The efficiency of stellar and AGN feedback is thus an important ingredient of galaxy evolution in the early universe. 

One of the most prominent effects of the feedback process is galactic outflows which have been extensively studied from local universe ($z\sim0$) to cosmic noon ($z\sim3$). Various feedback mechanism may drive outflowing gas with the bulk motions of a few hundreds to over one thousand km s$^{-1}$ that can be traced by broad emission lines \citep[e.g.,][]{Arribas+14,RodriguezdelPino+19,Swinbank+19} or absorption lines \citep[e.g.,][]{Heckman00,Rupke+05}. Several studies have also been conducted at $z\sim6$ using absorption lines \citep[][]{Sugahara+19} or emission lines with submillimeter observations \citep[][]{Herrera-Camus+21}. Local studies find outflows velocities are well correlated with SFR leading to the question of whether gas escape from the dark matter halo or return as fountains. Massive galaxies can drive high outflow velocity that is however significantly smaller than escape velocity \citep[e.g.,][]{Arribas+14}. Whether outflow can escape in low-mass galaxies is still debatable \citep[][]{Xu+22}. Another important question is related to outflow geometry. They may be biconical with small opening angle or more spherical with wide opening angle.

Recent JWST observations provide rich rest-frame optical spectroscopic data, via which outflows have been studied by several authors. For example, \cite{Tang+23}, \cite{Zhang+23}, and \cite{Carniani+23} (hereafter C23) report the detection of broad emission-line components in the H$\alpha$ and \Oiii~emission which traces the ionized phase of outflows. \cite{Zhang+23} also identify galaxies with spatially extended \Oiii~emission that do not broadly overlap with those with a broad emission-line, which can be explained by the duty-cycle from early outflows to late outflows. While simulations show outflows take place instantaneously instead of continuously \citep[e.g.,][]{Pandya+21}, no observational constraints on the duty-cycle of outflows have been given. 

In this study we search for outflow signatures in $3<z<9$ galaxies using the NIRSpec and NIRCam WFSS datasets obtained in multiple programs. We make the largest spectroscopic sample that is used to probe feedback efficiency and statistically reveal a picture of high-$z$ outflows. This paper is structured as follows: In Section \ref{dataset}, we describe our datasets and data reductions. In Section \ref{fit_outflows}, we explain how outflows signatures are identified. We present our results on outflow properties in Section \ref{results} and summarize our findings in Section \ref{summary}. Throughout the paper we adopt a cosmological model with $H_0=70~\mathrm{km~s^{-1}~Mpc^{-1}}$, $\Omega_\Lambda=0.7$, and $\Omega_{m}=0.3$.

\begin{figure*}[thb!]
\centering
\includegraphics[width=\linewidth]{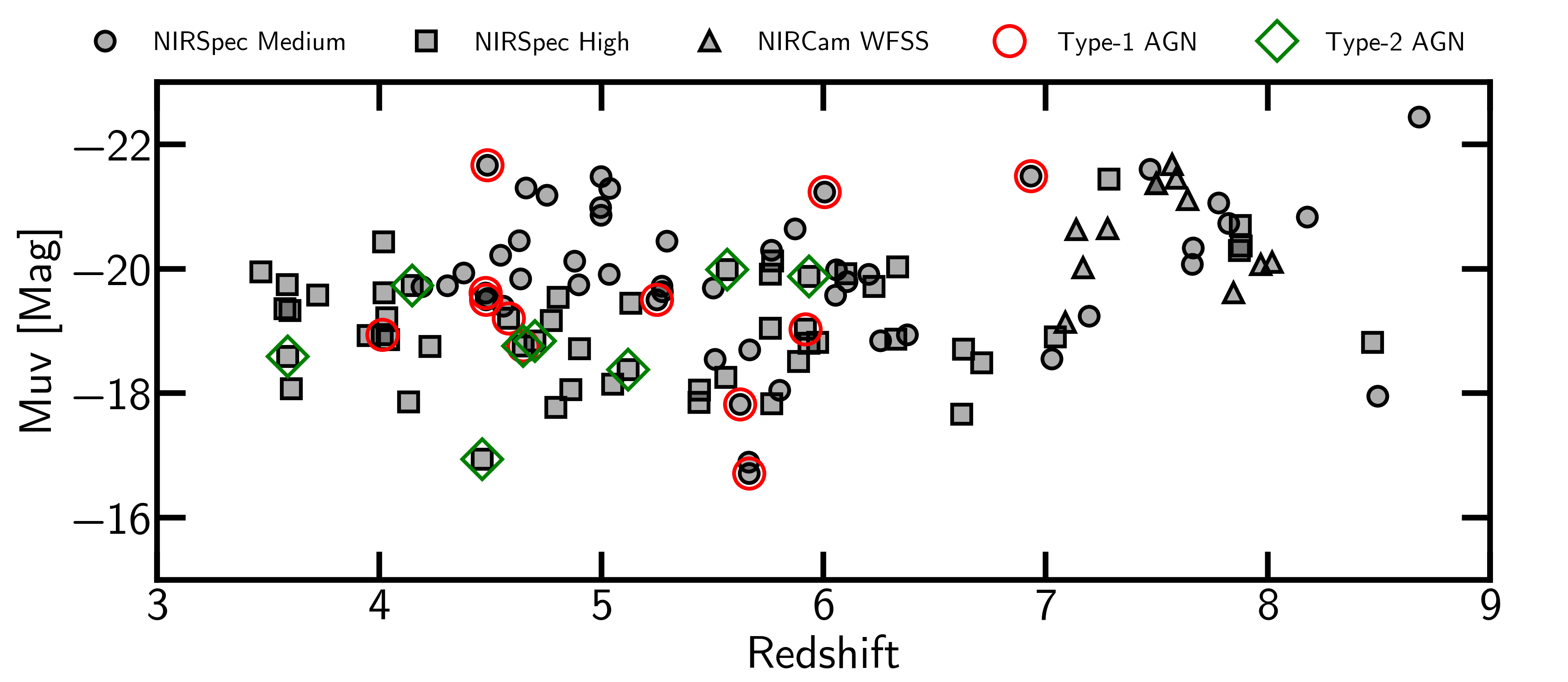}
\caption{
Redshift and $\Muv$ of the galaxy sample investigated in this study. Emission lines of H$\alpha$ and/or \Oiii$\lambda5007$ have been detected in the JWST spectroscopic data for each galaxy. The filled circles are galaxies observed with the setup of NIRSpec medium resolution taken from CEERS and ERO, while the squares are observed with the setup of NIRSpec high resolution taken from GLASS and JADES. The triangles are observed with NIRCam WFSS taken from FRESCO. The open circles denote the twelve Type-1 AGNs selected by \cite{Harikane+23b} and \cite{Maiolino+23c}. \noedit{The open diamonds are the Type-2 AGNs selected by \cite{Scholtz+23} from the JADES dataset}.
}
\label{fig:sample}
\end{figure*}

\begin{figure}[thb!]
\centering
\includegraphics[width=\linewidth]{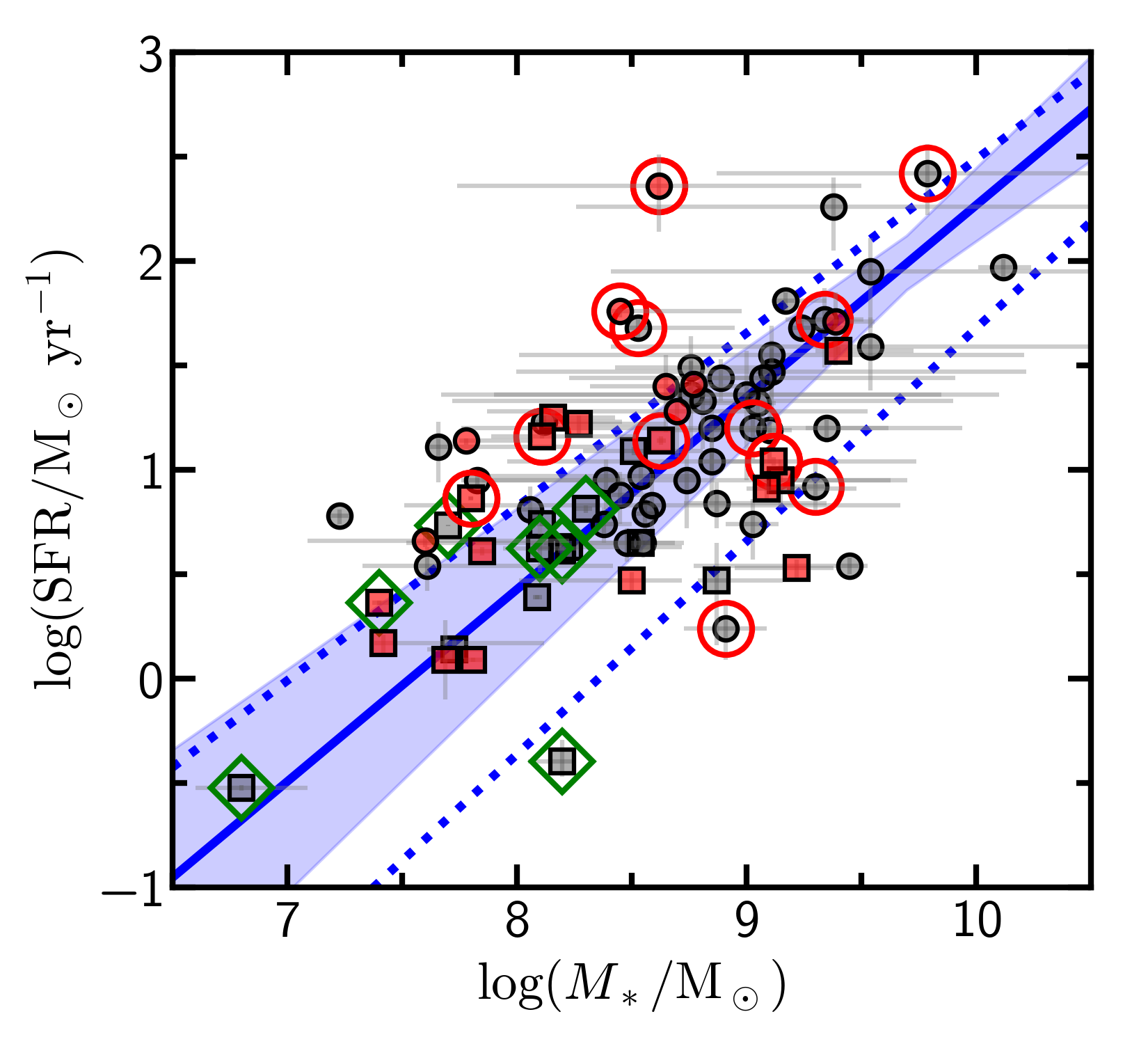}
\caption{
Same as Figure \ref{fig:sample} but for stellar masses and SFRs. Red filled symbols are galaxies with outflow signatures. We only include the galaxies whose stellar mass are derived from SED fitting. The blue lines are the star forming main sequences at $z\sim4$, 6, and 8 (from bottom to top) obtained by \cite{Santini+17}. The uncertainties at $z\sim6$ are indicated by the blue shaded region.
}
\label{fig:SFR_Mstar}
\end{figure}

\section{Observational Dataset and Galaxy Sample}
\label{dataset}

\subsection{NIRSpec}

We use the JWST/NIRSpec spectra reduced in \cite{Nakajima+23} and \cite{Bunker+23a} that are briefly described here. 
The data sets were obtained in the Early Release Observations (EROs; \citealt{Pontoppdan+22}) targeting the SMACS 0723 lensing cluster field (ERO-2736, PI: K. Pontoppidan), the Early Release Science (ERS) observations of GLASS (ERS-1324, PI: T. Treu; \citealt{Treu+22}), the Cosmic Evolution Early Release Science (CEERS; ERS-1345, PI: S. Finkelstein; \citealt{Finkelstein+23}, \citealt{ArrabalHaro+23}), and the JWST Advanced Deep Extragalactic Survey (JADES; GTO-1180, PI: D. Eisenstein; GTO-1210, PI: N. Luetzgendorf; \citealt{Eisenstein+23}, \citealt{Bunker+23a}).
The ERO data were taken in the medium resolution ($R\sim1000$) filter-grating pairs F170LP-G235M and F290LP-G395M covering the wavelength ranges of $1.7-3.1$ and $2.9-5.1$ $\mu$m, respectively.
The total exposure time of the ERO data is 4.86 hours for each filter-grating pair.
The GLASS data were taken with high resolution ($R\sim2700$) filter-grating pairs of F100LP-G140H, F170LP-G235H, and F290LP-G395H covering the wavelength ranges of $1.0-1.6$, $1.7-3.1$ and $2.9-5.1$ $\mu$m, respectively.
The total exposure time of the GLASS data is 4.9 hours for each filter-grating pair.
The CEERS data were taken with the Prism ($R\sim100$) that covers $0.6-5.3$ and medium-resolution filter-grating pairs of F100LP-G140M, F170LP-G235M, and F290LP-G395M covering the wavelength ranges of $1.0-1.6$, $1.7-3.1$ and $2.9-5.1$ $\mu$m, respectively.
The total exposure time of the CEERS data is 0.86 hours for each filter-grating pair.
The ERP, GLASS, and CEERS data were reduced with the JWST pipeline version 1.8.5 with the Calibration Reference Data System (CRDS) context file of {\tt jwst\_1064.pmap} with additional processes improving the flux calibration, noise estimate, and the composition.
Please see \cite{Nakajima+23} for details of the data reduction.
The JADES data were taken with the low-resolution prism and four filter-grating pairs (F070LP-G140M, F170LP-G235M, F290LP-G395M and F290LP-G395H). 
The total exposure time of the JADES data is 9.3 hours for the prism and 2.3 hours for each filter-grating pair.
Please see \cite{Bunker+23a} for details of the data reduction.

\subsection{NIRCam WFSS}
\label{wfss_reduc}

We use the data obtained in the First Reionization Epoch Spectroscopically Complete Observations (FRESCO; GO-1895, PI: P. Oesch; \citealt{Oesch+23}).
FRESCO took NIRCam imaging and slitless grism data in the two GOODS fields.
% copied from Oesch+
In each field, eight frames of grism data were taken at each of the eight pointing positions with a total exposure time of 2.0 hours at each position.
FRESCO used the F444W filter for the imaging and slitless grism data.
The F210M and F182M filters are used for the parallel imaging.
The grism data were taken with the GrismR configuration with a resolving power of $R\sim1600$. 
Here we describe the data reductions of the slitless grism data and the identification of \Oiii$\lambda\lambda4959,5007$ doublets.

We reduce the slitless grism data using a combination of the official pipeline and customized scripts that are motivated by the EIGER program \citep[][]{Kashino+23}.
We use the JWST pipeline version 1.9.6 with the Calibration Reference Data System (CRDS) context file of \texttt{jwst\_1106.pmap}.
We first process individual Level1 product using the assign wcs step from Spec2 to obtain the WCS solution for each grism image. 
We then process the data using Image2 for flat-fielding.
The large-scale sky background variations and the 1/f noise are removed by preforming a straight median-subtraction in each detector column (orthogonal to the dispersion direction).
The image after flat-fielding, background and 1/f noise subtraction are called the science image.
Following the methods in \cite{Kashino+23}, we apply median filtering using a boxcar-like kernel to each detector row of the science image which produces the image of continuum.
Because FRESCO uses the F444W filter instead of the F356W filter used by EIGER which targets on \Oiii~emitters at higher redshifts, we use a median filtering kernel with a length of 63 pixels and a central hole of 11 pixels that is larger than the one used by \cite{Kashino+23}.
The image of emission lines (hereafter emission image) is subsequently obtained by subtracting the continuum image from the science image.

We identify emission lines and remove contamination using the information from the the imaging data and the emission images.
Photometric redshifts can be good references to identify galaxies that have only one bright emission line within the wavelength coverage, such as H$\alpha$ \citep[e.g., ][]{Helton+23}.
In this study, we focus on the emission line doublets of \Oiii$\lambda\lambda4959,5007$ that can be selected without accurate photometric redshifts.
We first use \texttt{SExactor} \citep[][]{sextractor} on the NIRCam images taken with the F444W filter to create a parent catalog of photometric objects.
Similarly, we use \texttt{SExactor} on the emission images to make a catalog of emission lines.
For each photometric object and the wavelength range of $3.9-5.0~\mu\mathrm{m}$, we use \texttt{grismconf}\footnote{\url{https://github.com/npirzkal/GRISMCONF}} to calculate the positions (x, y) in the emission image where the corresponding spectrum locates.
We adopt the NIRCam grism model (V4)\footnote{\url{https://github.com/npirzkal/GRISM_NIRCAM}}, based on the commissioning observations.
We look for emission lines with an offset of 3 pixels in the detector column to the location of the spectrum.
We then search for emission line pairs with a wavelength separation of $380-480~\mathrm{\AA}$ that are possible \Oiii~doublets at $7<z<9$.
After selecting several candidates, we extract the 2d and 1d spectra and remove contaminations via visual inspections.
We perform a standard $\sigma$-clipping to combine the 8 frames to produce the coadd 2d and 1d spectra.
Finally we obtain twelve \Oiii~emitters in the FRESCO data as shown in Figures \ref{fig:FRESCO_2dspec}, \ref{fig:FRESCO_2dspec_2}, and Table \ref{tab:FRESCO_objects}.

\subsection{Sample and galaxy properties}
\label{galaxy_sample}

Outflow presenting as broad emission lines with FWHM$\gtrsim300~\mathrm{km~s^{-1}}$ can ideally be resolved by the Medium ($R\sim1000$) or High ($R\sim2700$) resolution of the NIRSpec spectra and the $R\sim1600$ NIRCam WFSS spectra.
We focus on strong rest-frame optical lines of of H$\alpha$ and \Oiii$\lambda5007$ that are good tracers of the ionized gas.
We study the objects with either emission line of H$\alpha$ or \Oiii~detected with a signal-to-noise ratio of S/N$>10$ in the NIRSpec data.
We fully exploit the wavelength coverage of NIRSpec data with Medium and High resolution and select 5, 48, 13, and 52 galaxies from ERO, CEERS, GLASS, and JADES, respectively.
We also include the 12 $z\sim8$ galaxies taken with the NIRCam WFSS data (see Section \ref{wfss_reduc}).
The final sample consists of 130 ($=5+48+13+52+12$) galaxies at $3<z<9$.
For the galaxies at $6<z<9$, only the \Oiii~lines are within the wavelength coverage.

The redshifts of our galaxies are determined by the central wavelenths of H$\alpha$ and/or \Oiii~lines.
For the galaxies taken from ERO, CEERS, and GLASS data, we adopt the rest-frame UV magnitude $\Muv$, stellar mass, SFR from the catalog made by \cite{Nakajima+23}.
We adopt the stellar mass and SFR reported by \Car~for 14 galaxies taken from JADES.
For the other 38($=52-14$) galaxies taken from JADES and 12 galaxies from FRESCO, we first obtain $\Muv$ from the photometric catalog of JADES \citep[][]{Eisenstein+23} and CANDELS \citep[][]{Grogin+11,Koekemoer+11}, respectively.
We derive the values of $\Muv$ assuming zero K-correction:
\begin{equation}
    \Muv =m+2.5\log(1+z )-5\log\left[\frac{d_\mathrm{L}(z)}{10~\mathrm{pc}}\right],
\end{equation}
where $m$ is the magnitude in the filter whose central wavelength is closest to the rest-frame wavelength of $1500~\mathrm{\AA}$ and $d_\mathrm{L}$ is the luminosity distance in the unit of parsecs.
We then estimate the SFR from the $\Muv$ values adopting the relation of \cite{Madau&Dickinson14} assuming the stellar initial mass function (IMF) of \cite{Chabrier03}.
We correct the SFR for the dust attenuation using an attenuation--UV slope ($\beta_\mathrm{UV}$) relation \citep{Meurer+99} and $\beta_\mathrm{UV}$--$\Muv$ relation at each redshift \citep{Bouwens+14}.
Assuming our galaxies locate at the star formation main sequence at $z\sim6$ determined by \cite{Santini+17}, we estimate the stellar masses from the SFRs.
We present the redshifts and $\Muv$ of all the 130 galaxies in Figure \ref{fig:sample} and show SFR as a function of stellar mass in Figure \ref{fig:SFR_Mstar} for those estimated from SED fitting.
{We measure a scatter of $\sim0.49$ dex around the $z\sim6$ main sequence for the stellar masses. This scatter and the uncertainties of $\Muv$ are propagated to the stellar masses that are not obtained by SED fitting.}
Although a linear fitting to the SFRs and stellar masses of our galaxies gives a flatter slope than the one determined by \cite{Santini+17}, it is probably due to selection bias since the limiting SFR is $\sim2.5~\mathrm{M_\odot~yr^{-1}}$ (converted from an H$\alpha$ flux of $3\times10^{-19}~\mathrm{erg~cm^{-2}~s^{-1}}$ at $z\sim6$ adopting the conversion factor of \citealt{Kennicutt+98}).
We find that the SFRs and stellar masses of our galaxies are broadly consistent with the star formation main sequence at $z\sim6$.

We also estimate the mass ($M_\mathrm{h}$) and circular velocity ($\vcir$) of the dark matter halos that host the galaxies.
The dark matter halo mass is derived from the $M_\mathrm{h}-\Muv$ relation at each redshift \citep[][]{Harikane+22}.
Then we adopt the equations in \cite{Mo&White02} for $v_\mathrm{cir}$:
\begin{equation}
    v_\mathrm{cir}=\left(\frac{GM_\mathrm{h}}{r_\mathrm{h}}\right)^{1/2},
\end{equation}
\begin{equation}
    r_\mathrm{h}=\left(\frac{GM_\mathrm{h}}{100\Omega_\mathrm{m}H_0^2}\right)^{1/3}(1+z)^{-1},
\end{equation}
where $r_\mathrm{h}$ is the halo radius defined by an overdensity of 200 times the cosmic average density.

We notice that several galaxies in our sample are reported as Type-1 AGN seleced by broad H$\alpha$ lines \citep[][]{Harikane+23b,Maiolino+23c}.
Although there are prominent broad emission-line components originated from the broad line region (BLR), the outflows can be detected via detailed modeling of emission lines (see Section \ref{fit_AGN}).
\noedit{For the galaxies from JADES, we also compare the results with the Type-2 AGNs from \cite{Scholtz+23} that do not present clear broad components from BLR. Although \cite{Scholtz+23} select Type-2 AGNs with multiple high ionization lines, it is considered difficult to distinguish high-z AGNs with star forming galaxies based on line ratio diagnostics. Therefore, this sample of Type-2 AGN may not be complete. Type-1 and Type-2 AGNs are highlighted with red circles and green diamonds, respectively, in Figure \ref{fig:sample} and \ref{fig:SFR_Mstar}.}

\begin{figure*}[t!]
\centering
\includegraphics[width=0.33\linewidth]{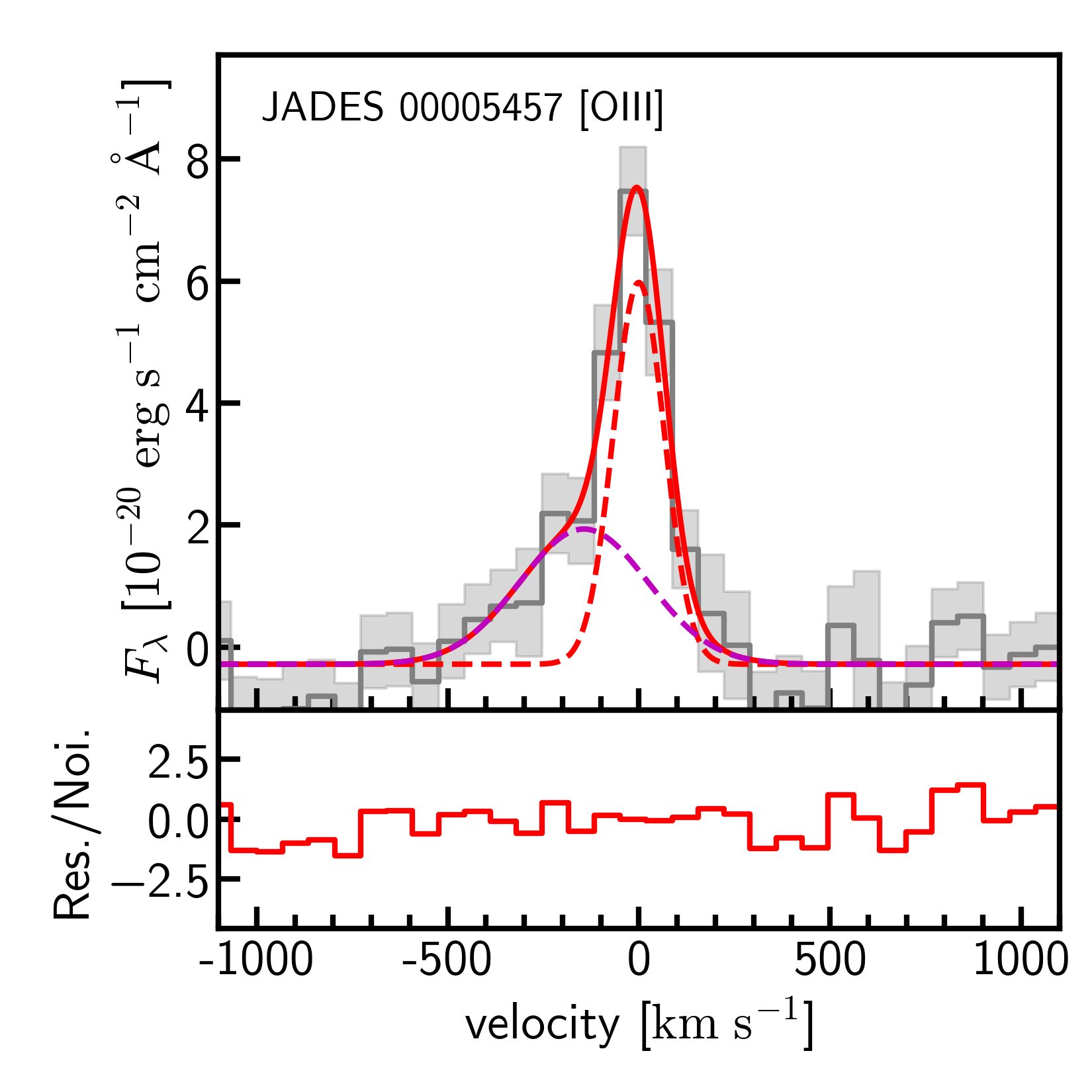}%
\includegraphics[width=0.33\linewidth]{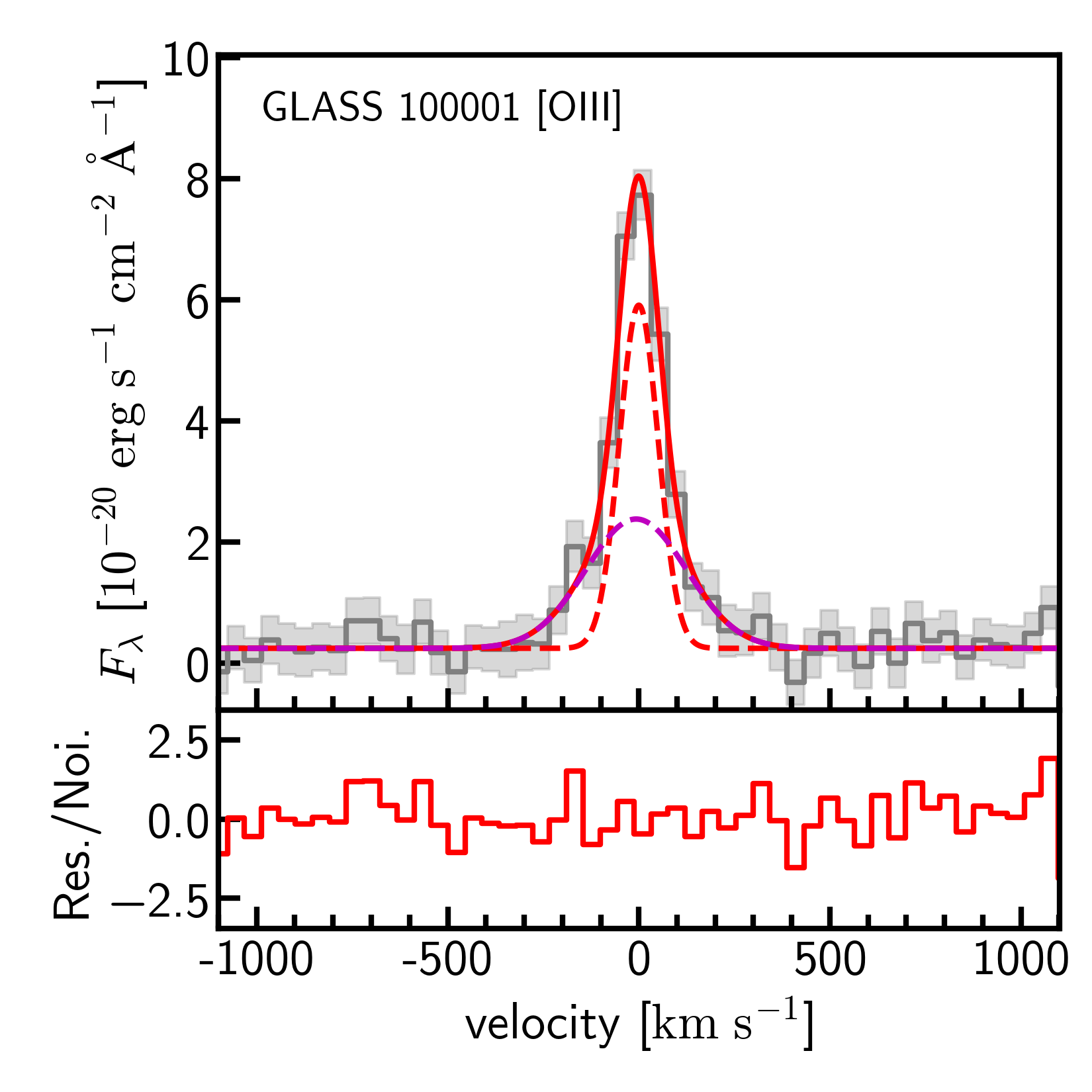} \\
\includegraphics[width=0.33\linewidth]{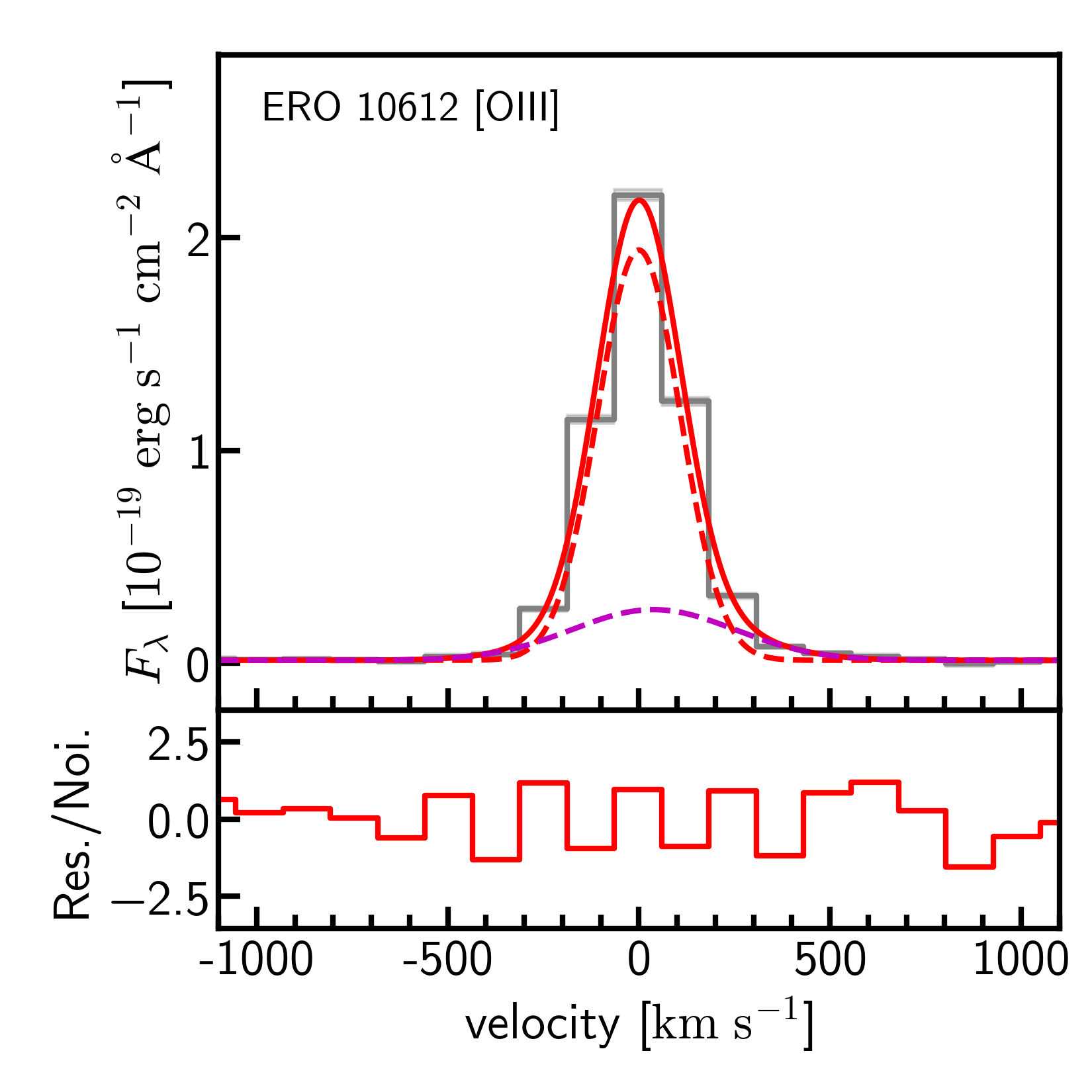}%
\includegraphics[width=0.33\linewidth]{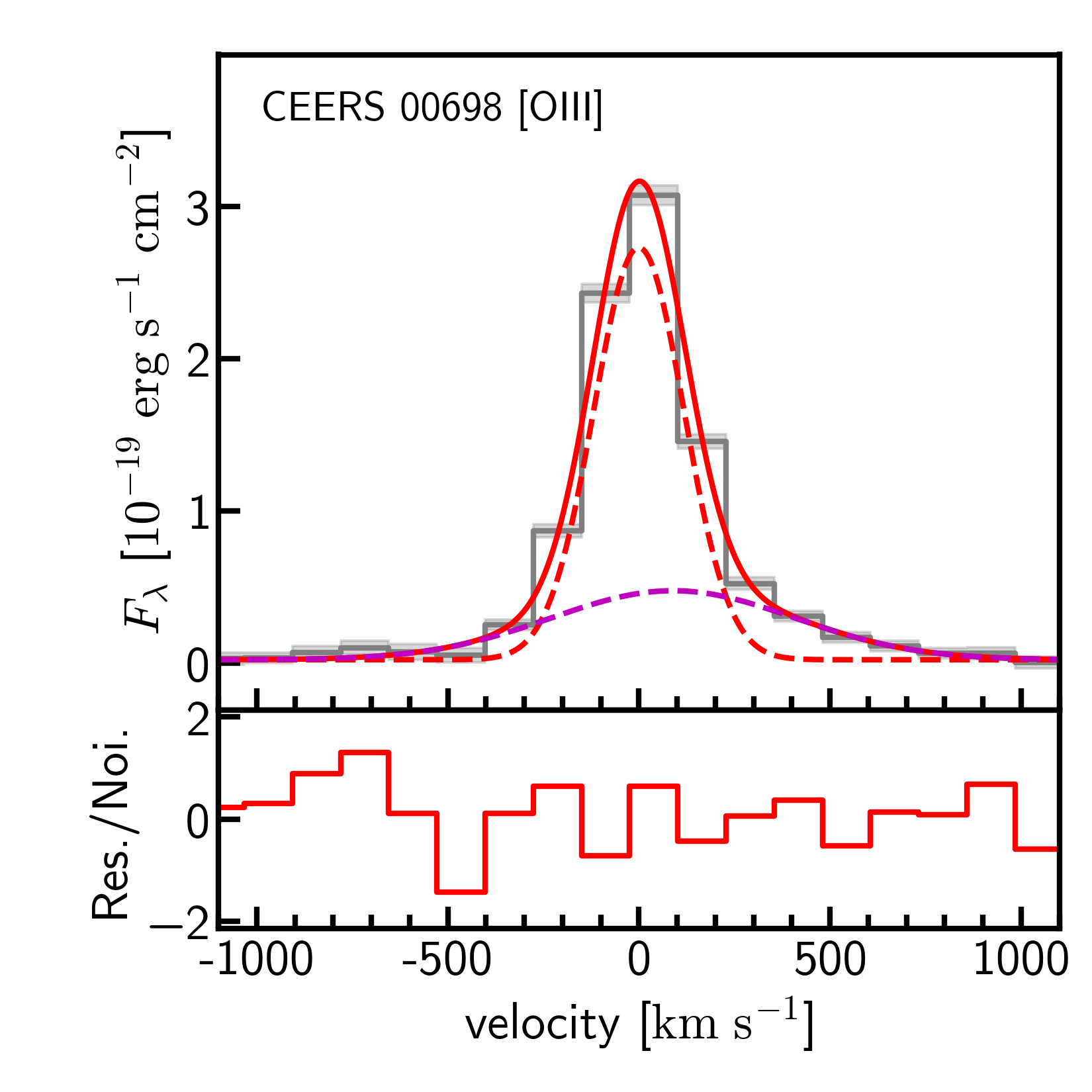}%
\includegraphics[width=0.33\linewidth]{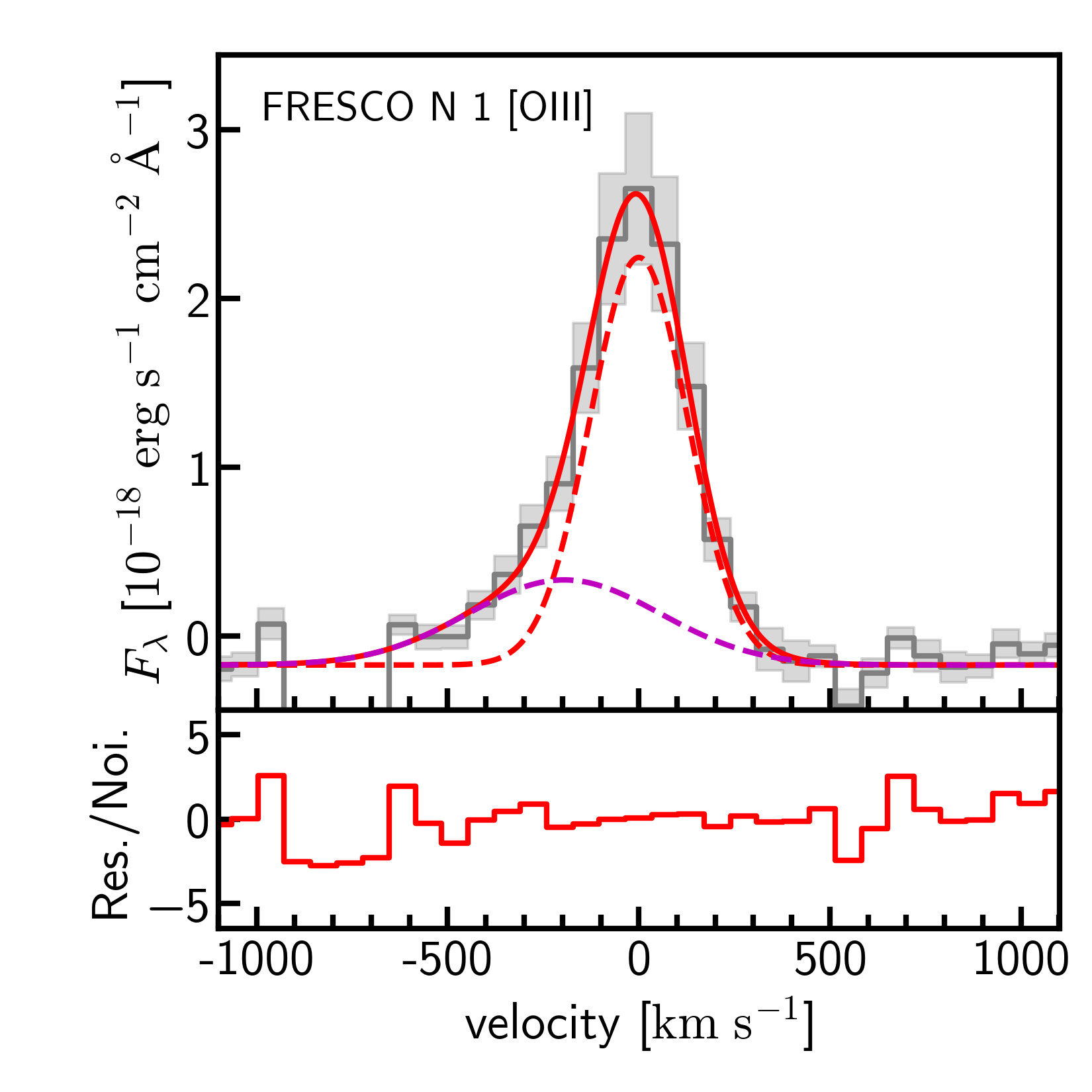}
\caption{
Example spectra fitted with two Gaussian components. The top panels show the spectra (grey histogram) with the associated uncertainties (grey shades), narrow component (red dashed line), the outflow component (magenta dashed line), and the overall shape of the best-fit profile (red solid line). The bottom panels show the residuals of the best-fit profile normalized by the uncertainties of flux density.
}
\label{fig:fitting}
\end{figure*}

\begin{figure*}[t!]
\centering
\includegraphics[width=\linewidth]{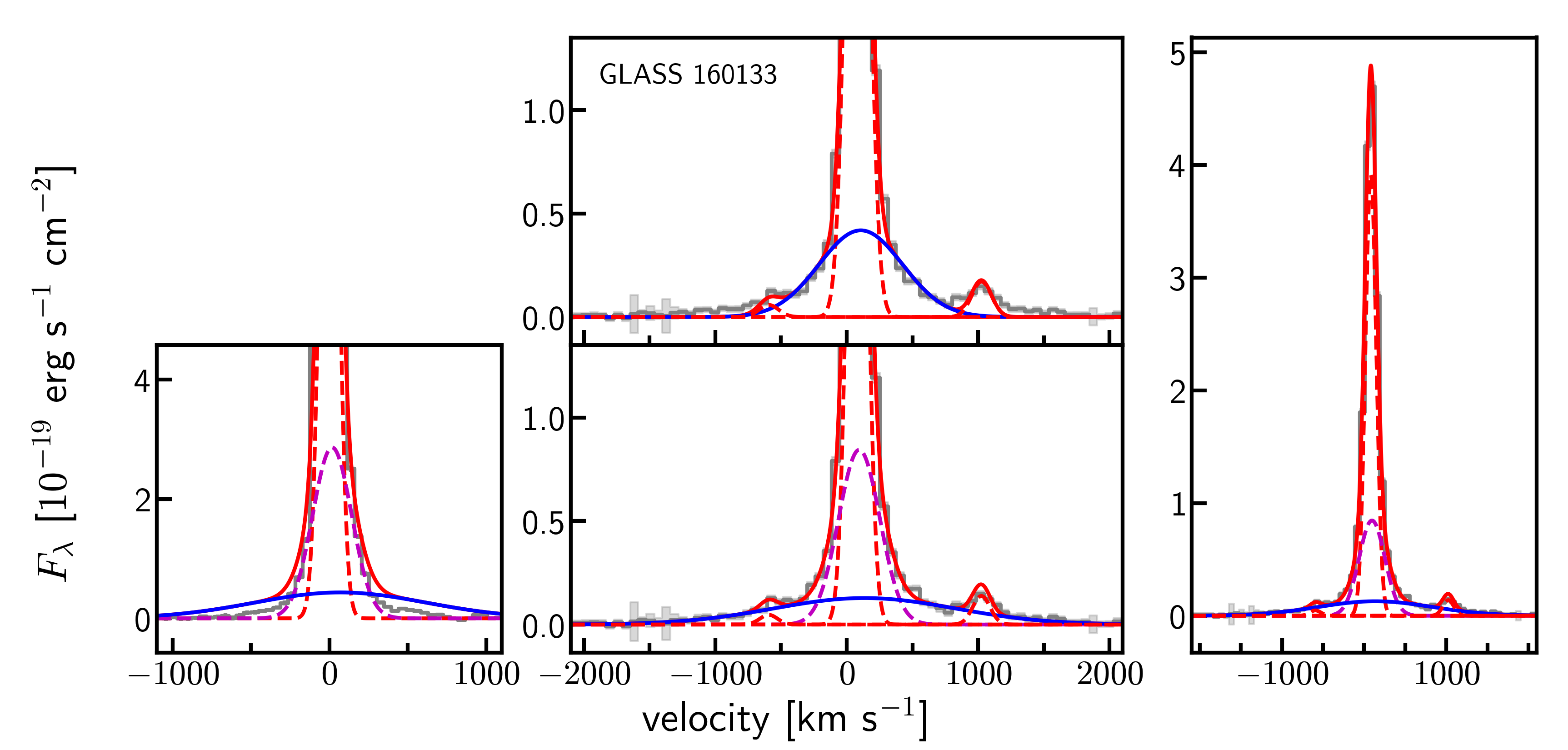}
\caption{
Example spectrum of a Type-1 AGN. The H$\alpha$ lines in the right panel are fitted with three Gaussian components (narrow component, outflow component, and BLR component). The middle panels present zoom-in of the H$\alpha$ line that are fitted with two (three) Gaussian components in the top (bottom) middle panel. The left panel shows the spectrum of the \Oiii~line. The three-component model is scaled and overplotted on the \Oiii~line.
}
\label{fig:fiting_AGN}
\end{figure*}

\section{Identification of Outflows}
\label{fit_outflows}

\subsection{outflows in star forming galaxies}

We search for the signatures of outflows by fitting the emission lines of \Oiii~and H$\alpha$ with two Gaussian components:
\begin{align}
F_\lambda(\lambda) = &A_\mathrm{n}\exp\left(-\frac{(\lambda - \lambda_\mathrm{n})^2}{2\sigma_\mathrm{n}^2}\right) + \\ \notag
&A_\mathrm{b}\exp\left(-\frac{(\lambda - \lambda_\mathrm{b})^2}{2\sigma_\mathrm{b}^2}\right) + C,
\label{eq:two_comp_model}
\end{align}
where $A_\mathrm{n}$, $\lambda_\mathrm{n}$, and $\sigma_\mathrm{n}$ are the amplitude, central wavelength, and line width of the narrow component, respectively.
The outflow component is characterized by $A_\mathrm{b}$, $\lambda_\mathrm{b}$, and $\sigma_\mathrm{b}$ in the same manner as the narrow component.
{A continuum level $C$ is also included as a free parameter.}
We perform a least $\chi^2$ fitting using a customized python script and the Levenberg-Marquardt algorithm implemented by the \texttt{scipy} package.
We do not fit the [{\sc Nii}] doublets simultaneously because the separation between H$\alpha$ and [{\sc Nii}]$\lambda6563$ is about $700~\mathrm{km~s^{-1}}$ that is 2--3 times larger than the typical line width of outflow component.

We look for a well-fitted and statically significant broad component to identify galaxies with outflows.
The broad component is required to have 1) $\sigma_\mathrm{b}>2\sigma_\mathrm{n}$ and 2) S/N $>$ 3.
The two-component models usually have smaller residuals and reduced $\chi^2$ values than those of single-component models due to the increase of free parameters.
Statistical indicators such as Akaike information criteria $\mathrm{AIC}=2n-\log(L)$ and Bayesian information criteria $\mathrm{BIC}=kn-\log(L)$ are often used to test whether a secondary Gaussian component is needed, where $n$ is the number of free parameters, $k$ is the number of data points, and $L$ is the likelihood of the best-fit model.
Recent studies such as \Car~also use the Bayesian factor $p(M|\theta)$ that is equivalent to BIC at low-order approximation: $p(M|\theta)\approx\exp(-\mathrm{BIC}/2)p(M)$.
We use $\Delta \mathrm{AIC=AIC(two~components) - AIC(one~component)}<-2$ to select statistically significant broad component because AIC has smaller penalty for the number of free parameters than BIC, which may include weaker outflow signatures.
When the evidence for secondary Gaussian component is strong (e.g., $\Delta \mathrm{AIC}<-10$), the results of AIC and BIC should be consistent.
We also visually inspect the fitting to rule out fake outflow signatures given by the noises.

We detect 8, 20, and 2 for the observational setup of NIRSpec medium resolution, NIRSpec high resolution, and NIRCam WFSS, respectively.
In Section \ref{rate}, we discuss the incidence rate of outflow in high-$z$ galaxies in details.

\subsection{outflows in AGNs}
\label{fit_AGN}

For 12 Type-1 AGNs from \cite{Harikane+23b} and \cite{Maiolino+23c}, we fit the H$\alpha$ lines with three Gaussian components including two broad components from BLR and outflow:
\begin{align}
F_\lambda(\lambda) = &A_\mathrm{n}\exp\left(-\frac{(\lambda - \lambda_\mathrm{n})^2}{2\sigma_\mathrm{n}^2}\right) + \\ \notag
&A_\mathrm{b}\exp\left(-\frac{(\lambda - \lambda_\mathrm{b})^2}{2\sigma_\mathrm{b}^2}\right) + \\ \notag
&A_\mathrm{BLR}\exp\left(-\frac{(\lambda - \lambda_\mathrm{BLR})^2}{2\sigma_\mathrm{BLR}^2}\right) + C,
\end{align}
where the BLR component is characterized by $A_\mathrm{BLR}$, $\lambda_\mathrm{BLR}$, and $\sigma_\mathrm{BLR}$ in the same manner as the narrow component.
We simultaneously fit the [{\sc Nii}] doublets with the velocity shift and velocity dispersion fixed to those of the narrow component.

We identify a significant third component in CEERS 01244, GLASS 150029, GLASS 160133, and JADES 00008083 by comparing the AIC values between the two-component model and the three-component model. However, the complex profile consisting of multiple broad components may originate from dual BHs \citep{Maiolino+23c} or the structure of BLR \citep{Kollatschny+13} instead of outflows. For GLASS 150029 and GLASS 160133 whose \Oiii~lines are within the wavelength coverage of NIRSpec data, we scale the triple-Gaussian profile by the flux ratio between \Oiii~and H$\alpha$ and overplot the profile on the left panel of Figure \ref{fig:fiting_AGN} as a sanity check. As we expect, the BLR component cannot been seen in the forbidden \Oiii~line while the outflow component can well explain the broad wings between the narrow component and emission line profile, which strongly suggests the existence of outflow components.

\newpage
\begin{figure*}[tbh!]
\centering
\includegraphics[width=\linewidth]{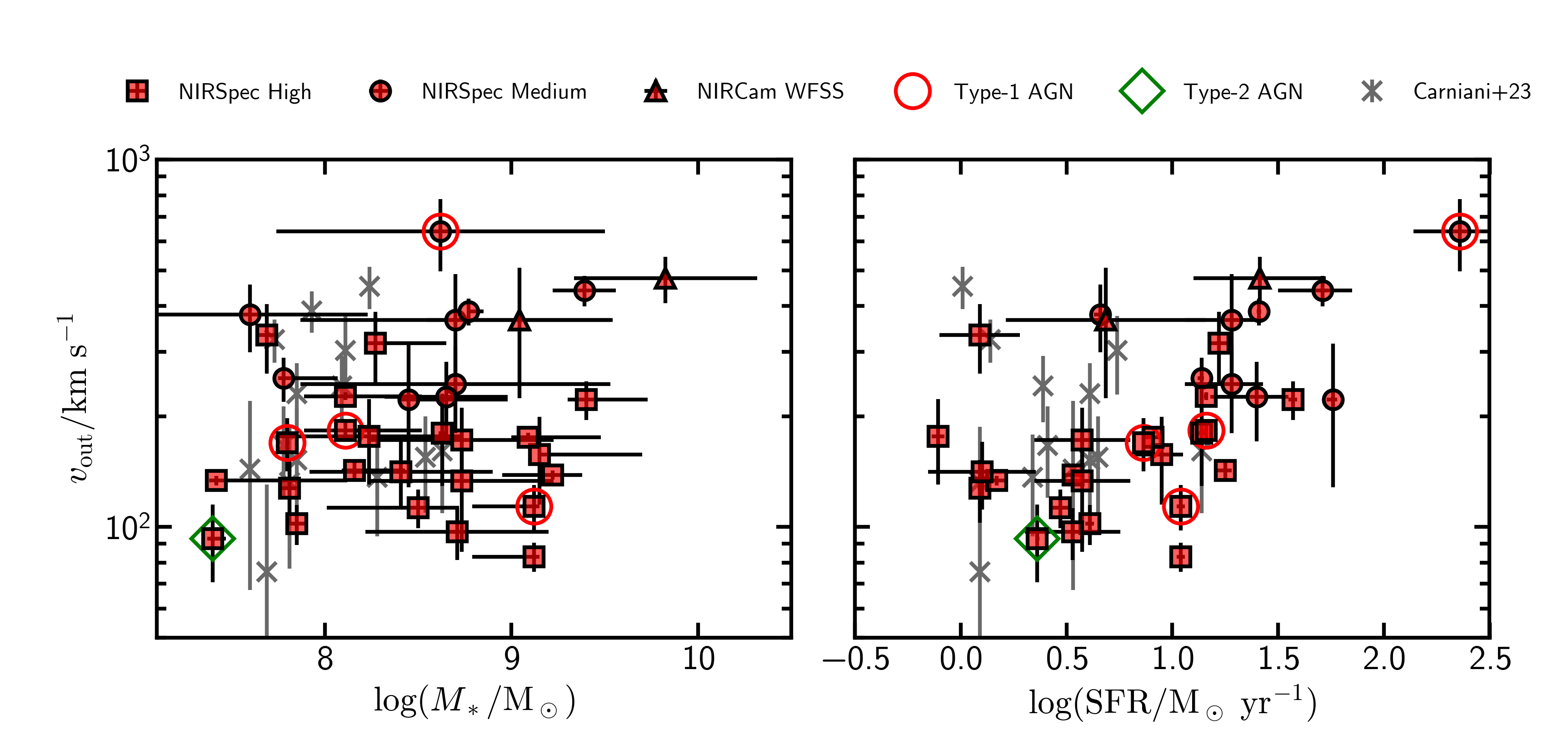}
\caption{
Outflow velocity as a function of stellar mass and SFR. \noedit{The red symbols summarize our measurements of outflow velocities using different instrument setups. The grey crosses are measured by \Car\ based on the high resolution spectra of JADES.}
}
\label{fig:outflow_velocity}
\end{figure*}

\begin{figure*}[tbh!]
\centering
\includegraphics[width=\linewidth]{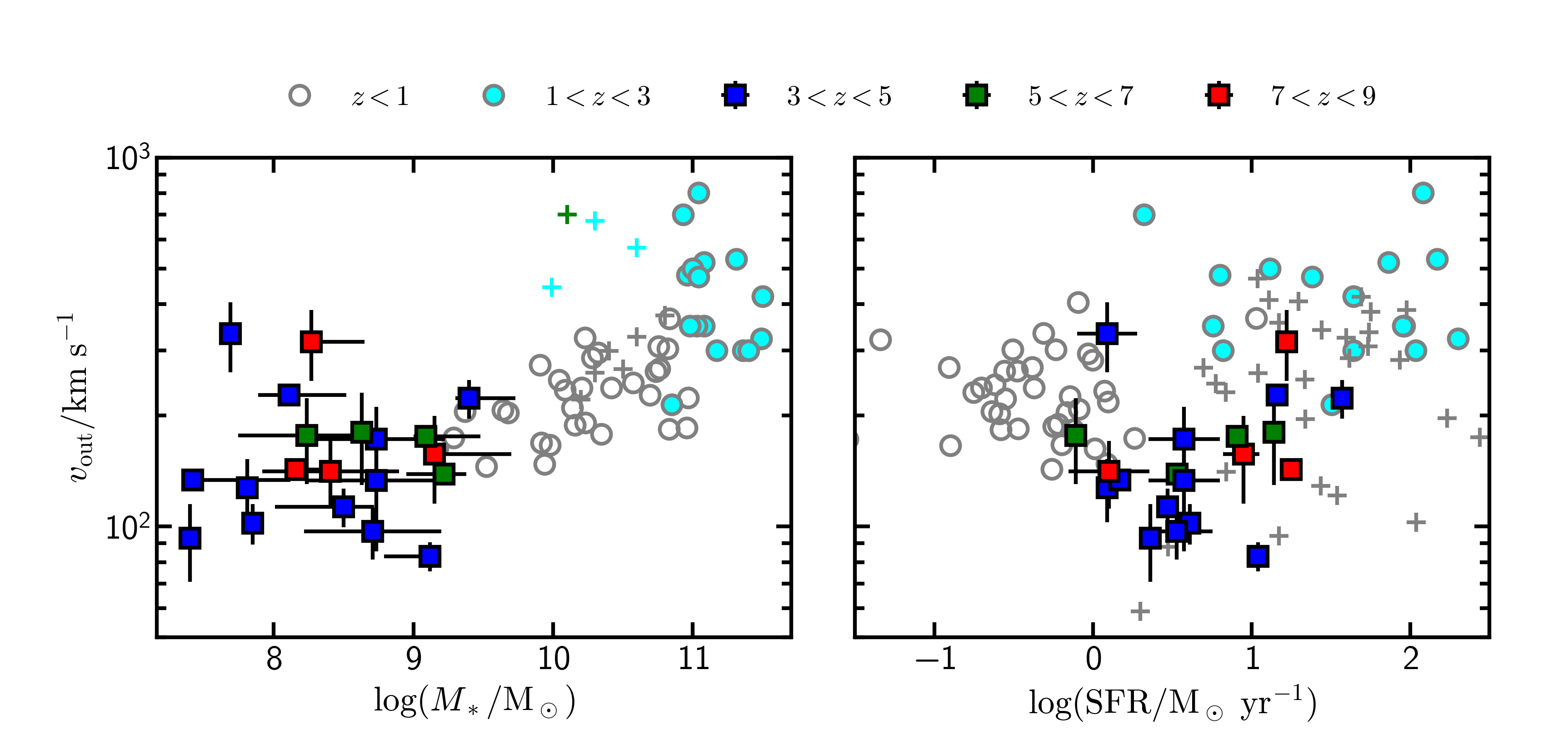}
\caption{
\noedit{Same as Figure \ref{fig:outflow_velocity} but including only velocities measured with NIRSpec high resolution spectra. The grey, cyan, blue, green, and red colors indicate galaxies at increasing redshifts. Open circles and cyan circles are $z\sim0$ galaxies whose outflow velocities are measured with emission lines by \citep{RodriguezdelPino+19} and $z\sim2$ \citep{Genzel+14}, respectively. We also include the outflow velocities measured with absorption lines (crosses) for the discussions of mass dependence \citep{Sugahara+17,Sugahara+19} and SFR dependence from \citep{Heckman00}.}
}
\label{fig:outflow_velocity_redshift}
\end{figure*}

% \newpage

\section{Results and Discussions}
\label{results}

\subsection{outflow velocity}
We study the maximum outflow velocity in the line-of-sight direction defined as:
\begin{equation}
    v_\mathrm{out}=|\Delta v| + \mathrm{FWHM_b}/2,
    \label{eq:vout}
\end{equation}
where $\Delta v$ is the velocity shift between the narrow and outflow component.
This definition has been previously used by several studies showing good correlations with SFR \citep[][]{Arribas+14} or stellar mass \citep[][]{RodriguezdelPino+19} for $z\sim0$ galaxies.
The FWHM of the outflow component is calculated from $\sigma_\mathrm{b}$ in Equation \ref{eq:two_comp_model} after subtracting the instrumental broadening, $\sigma_\mathrm{inst}$.
For the NIRSpec data, we adopt the dispersion of the line spread functions measured by \cite{Isobe+23b}, who calibrate the LSF for different filter-grating pairs and wavelengths.
For NIRCam WFSS data, we adopt $\sigma_\mathrm{inst}\sim80~\mathrm{km~s^{-1}}$ consistent with the resolution of NIRCam grisms.
We obtain $\vout\sim80-500~\mathrm{km~s^{-1}}$ with an average value of $190~\mathrm{km~s^{-1}}$.
For the different instrument setups of NIRSpec high resolution, NIRSpec medium resolution, and NIRCam WFSS, the average $\vout$ values are $130$, $300$, and $320~\mathrm{km~s^{-1}}$, respectively.
\noedit{The results are presented in Table \ref{tab:outflow} and Figure \ref{fig:outflow_velocity}.
Our measurements are consistent with those of galaxies taken from JADES studied by \Car. }
On average the outflow velocities derived from NIRSpec medium resolution and NIRCam WFSS are larger than those from NIRSpec high resolution because relatively slower outflows cannot be resolved by the lower resolutions. \noedit{We cannot thoroughly discuss the outflow velocities based on different spectral resolutions. Although the NIRSpec high resolution spectra can be convolved with the LSF to have consistent spectral resolution as the medium resolution spectra, the detection of outflows would get significantly weaker and biased towards fast outflows. We mainly consider the high resolution spectra in the following discussions. The galaxies with outflows detected with NIRSpec medium resolution and NIRCam WFSS spectra can be potential targets for follow-up observations in the future.}

In Figure \ref{fig:outflow_velocity_redshift}, we \noedit{discuss the outflow velocity as a function of stellar mass and SFR in different redshift ranges.
The sample size at each redshift ranges is not large enough to draw robust conclusions on the correlation between outflow velocities and stellar mass (SFR).
On the other hand, outflow velocities evolve weakly in the redshift range of $3<z<9$.
We also include previous measurements of outflow velocities from $z\sim0$ and $1<z<3$ using emission lines \citep{RodriguezdelPino+19,Genzel+14}.
Because high-z galaxies are less massive, our results probe the low-mass regime of $10^7-10^9~\mathrm{M_\odot}$ that is rarely investigated by previous studies.
We find a marginal correlation between $\vout$ and stellar mass across different redshifts, while the correlation between $\vout$ and SFR shows a more complicated trend.
In the right panel of Figure \ref{fig:outflow_velocity_redshift}, our results fall below the increasing trend between $\vout$ and SFR for $0<z<3$ galaxies.
This can be attributed to the low stellar masses of our galaxies, suggestive of that outflows may be strongly governed by the gravitational potential \citep{Concas+17,Sugahara+17,Sugahara+19}.
We also compare our outflow velocities to those measured from absorption lines.
Although the definition of $\vout$ may be different, similar trends can be identified.
% \cite{Sugahara+19} find faster outflows for a fixed stellar mass, which needs to be further tested with larger galaxy samples at different redshifts.
}
% In Figure \ref{fig:outflow_velocity_redshift}, we check the redshift dependence of $\vout$--stellar mass and SFR relations. We divide the galaxy sample with NIRSpec high resolution into three redshift bins. Although with relatively small sample size, .%
%{Note that we recalculate the outflow velocity based on their fitting results and Equation \ref{eq:vout} to compare $\vout$ in a consistent manner.}

In Figure \ref{fig:outflow_velocity}, we highlight the four Type-1 AGNs with red circles \noedit{and a Type-2 AGN with green diamond}. Interestingly, 3 out of the 4 Type-1 AGNs have similar outflow velocities as those of star forming galaxies. \noedit{The only Type-2 AGN with outflow signatures have $\vout\sim100~\mathrm{km~s^{-1}}$}. CEERS 01244 may host fast outflows with $\vout\gtrsim500~\mathrm{km~s^{-1}}$ driven by AGN activity or intensive star formation. However, the origin of the broad H$\alpha$ component is debatable due to the lack of \Oiii~emission. In general, there is no evidence of fast outflows powered by AGN activity in our high-$z$ galaxy sample. {It appears that outflow velocity is independent of Type-1 AGN activity, which may suggest the dominated role of supernovae and massive stars in driving outflows. However, significant contribution from AGNs is plausible if AGN activity is actually ubiquitous in high-$z$ galaxies \citep[e.g.,][]{Greene+24} while most of them are missed by the broad emission line selection.}

\begin{figure}[ht!]
    \centering
    \includegraphics[width=\linewidth]{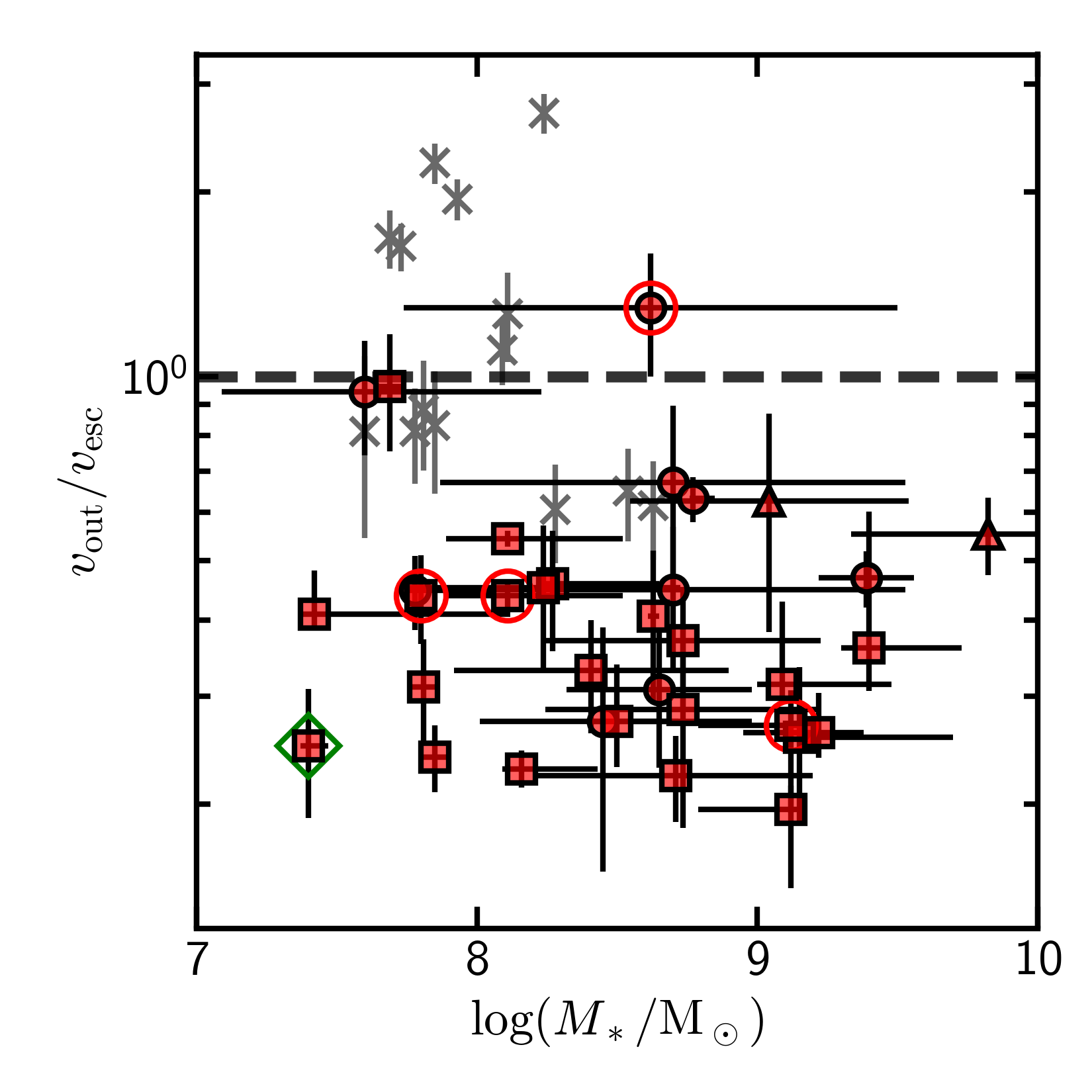}
    \caption{
    The ratio between $\vout$ and the escape velocity as a function of stellar mass. The symbols are the same as those in Figure \ref{fig:outflow_velocity}.
    }
    \label{fig:outflow_escape}
\end{figure}    

We test whether the ionized outflows have enough velocity to escape from the gravitational potential well by comparing the outflow velocity to the escape velocity $v_\mathrm{esc}$.
Following \cite{Heckman00}, we estimate the outflow velocity from the equation below:
\begin{equation}
    v_\mathrm{esc}=v_\mathrm{cir}\{2[1+\mathrm{ln}(r_\mathrm{max}/r)]\}^{\frac{1}{2}},
\end{equation}
where $r$ denotes the radius to evaluate $v_\mathrm{esc}$ usually approximated by the half-light radius.
The value of $r_\mathrm{max}$ is the halo radius calculated in Section \ref{galaxy_sample}.
% Note that the drag force is ignored in this estimation.
An approximation of $v_\mathrm{esc}=3v_\mathrm{cir}$ can be adopted because the escape velocity is not sensitive to the value of $r_\mathrm{max}/r$ in the range of $10-100$ \citep{Veilleux+05}.
\Car~use a software named \texttt{galpy} to evaluate $v_\mathrm{esc}$ at $z=0$ given by the gravitational potential of dark matter and galactic disk that can be described by the Navarro-Frenk-White (NFW) profile \citep[][]{Navarro+96} and the Hernquist profile \citep[][]{Hernquist90}, respectively.  
{We follow the methods in \Car~and find $3\vcir$ is a robust approximation of escape velocity comparing to the uncertainties on outflow velocity.}
In Figure \ref{fig:outflow_escape}, we show the ratio of $\vout/v_\mathrm{esc}$ as a function of stellar mass.
The ratios derived by \Car\ are displayed with grey crosses.
The difference between our results and \Car\ might be given by the techniques applied for spectra fitting and outflow detections.
With a sample size of 30 galaxies with outflows, we find almost all of the outflows at $3<z<9$ can not escape from the gravitational but return as fountains.

\begin{figure}[ht!]
\centering
\includegraphics[width=\linewidth]{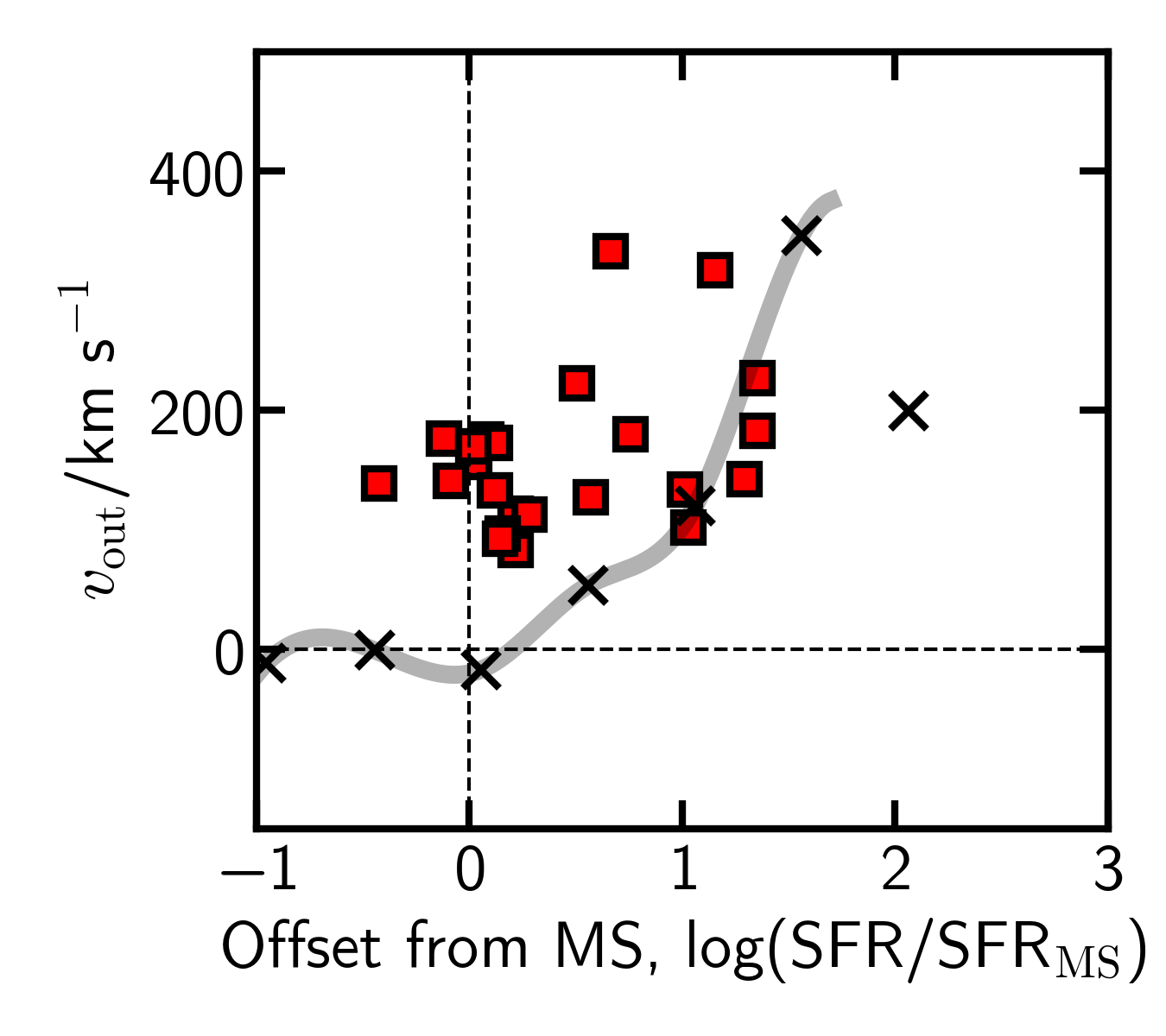}
\caption{
Outflow velocity as a function of offset from main sequence. Outflow velocities increase when galaxies have higher SFR than main sequence, which agrees with the trend observed in the composite spectra of $z\sim1$ galaxies shown with the black crosses \citep{Swinbank+19}. The grey line is an interpolation to the black crosses to highlight the increasing trend.
}
\label{fig:outflow_MS}
\end{figure}

\subsection{suppression of star formation}

\noedit{Stellar and AGN feedback are considered the main driver of star formation suppression \citep{Veilleux+05,Cicone+16,Concas+17} which determines the shape of star formation main sequence (MS).
\cite{Cicone+16} identify signatures of outflows only in galaxies above the MS. \cite{Swinbank+19} show that outflows are faster if galaxies are offset from the MS with larger SFR based on composite spectra of galaxies at $z\sim1$. The results from \cite{Swinbank+19} are presented by the crosses and the grey line (interpolated for presentation purpose) in Figure \ref{fig:outflow_MS}, which agrees with the physical picture that outflows strongly suppress the star formation for galaxies in a bursty phase. We also calculate the offsets of our galaxies from the MS using the best-fit equation to MS at different redshift obtained by \cite{Popesso+23}. Although the absolute values cannot be directly compared to those of \cite{Swinbank+19} due to different definitions of outflow velocities, our results in Figure \ref{fig:outflow_MS} suggest a similar increasing trend with MS offset albeit with a large scatter. The incidence of outflows are also more likely to be identified in galaxies above the MS suggestive of contribution to star formation suppression.
We note again that our galaxies are in general less massive, younger and more bursty than those studied by previous low-z studies.
The sample size at each redshift is also limited.
A more thorough galaxy sample is needed to conclude the relation between outflows and suppression on star formation at high z and to rule out the biases caused by sample selection and instrumental effects.}

\begin{figure*}[t!]
    \centering
    \includegraphics[width=0.75\linewidth]{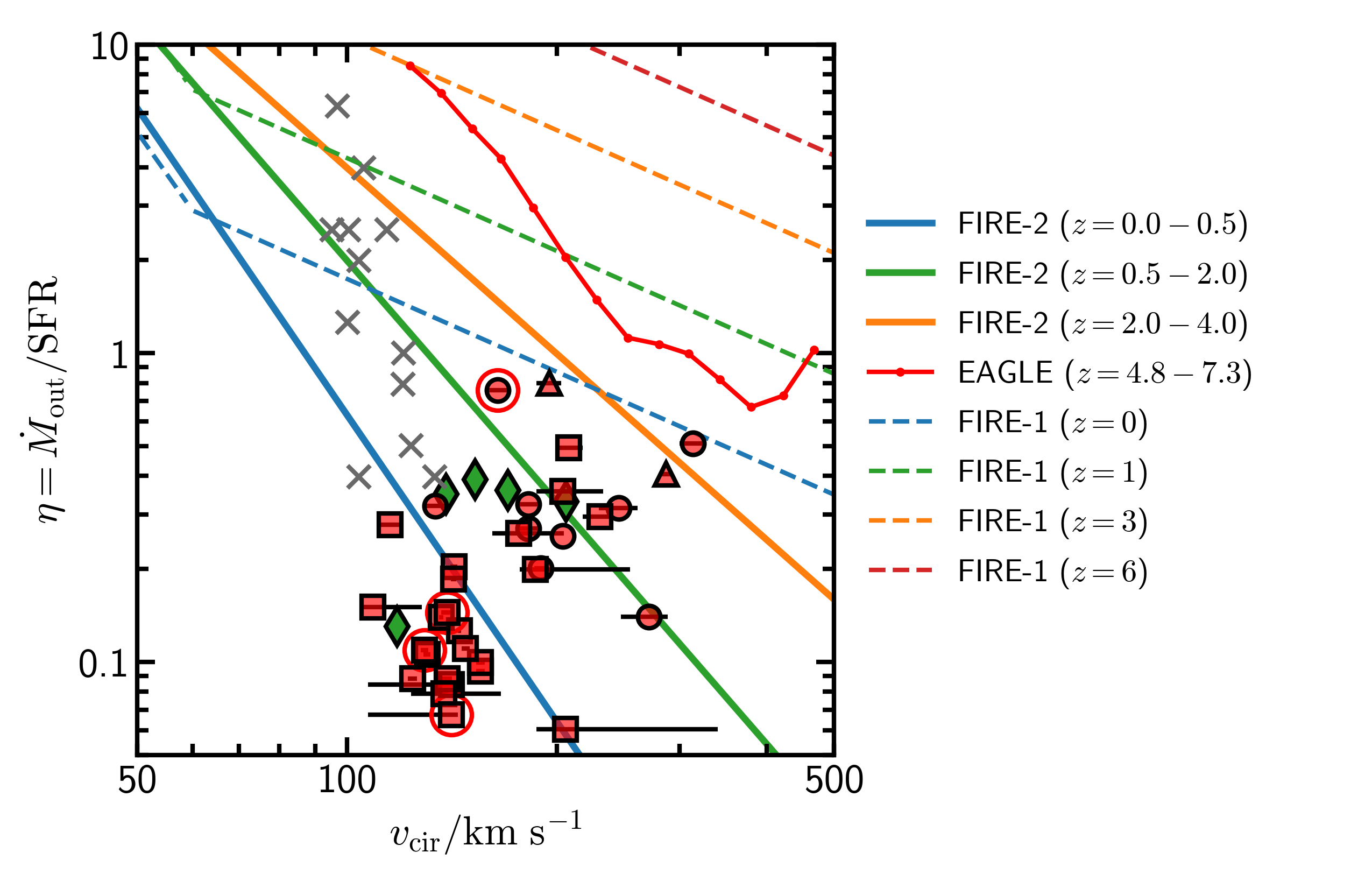}
    \caption{
    Mass loading factor $\eta$ as a function of $\vcir$. The symbols are the same as those in Figure \ref{fig:outflow_velocity}. We overplot the simulation predictions from FIRE-1 \citep[][]{Muratov+15}, FIRE-2 \citep[][]{Pandya+21}, and EAGLE \citep[][]{Mitchell+20}. Our observational results prefer inefficient feedback at high-$z$ with weak redshift dependence.
    }
    \label{fig:eta_vcir}
\end{figure*}

\subsection{feedback efficiency}
We estimate the mass outflow rate and the ratio of mass outflow rate to SFR known as the mass loading factor:
\begin{equation}
\label{eq:eta}
    \eta = \frac{\dot{M}_\mathrm{out}}{\mathrm{SFR}},
\end{equation}
where $\dot{M}_\mathrm{out}$ is the mass outflow rate.
Assuming the outflowing gas fills an ionized cone with a radius of $r_\mathrm{out}$,
The outflow mass can be estimated from the luminosity of the broad H$\alpha$ lines:
\begin{equation}
\label{eq:Mout}
    M_\mathrm{out}=\frac{1.36m_\mathrm{p}
    L_\mathrm{H\alpha,b}}{\gamma_\mathrm{H\alpha}n_\mathrm{e}},
\end{equation}
where $L_\mathrm{H\alpha,b}$ is the extinction-corrected H$\alpha$ luminosity of the broad component.
The quantity $m_\mathrm{p}$ is the atomic mass of hydrogen and $n_\mathrm{e}$ is the electron density in the outflowing gas.
For the volume emissivity of H$\alpha$ ($\gamma_\mathrm{H\alpha}$), we use $\gamma_\mathrm{H\alpha}=3.56\times10^{-25}~\mathrm{erg~cm^{-3}~s^{-1}}$, assuming the case B recombination and an electron temperature of $T_\mathrm{e}=10^4~\mathrm{K}$.
We calculate the mass outflow rate by dividing the outflow mass and the dynamical timescale, $r_\mathrm{out}/v_\mathrm{out}$:
\begin{equation}
\label{eq:dotMout}
    \dot{M}_\mathrm{out}=M_\mathrm{out}\frac{v_\mathrm{out}}{r_\mathrm{out}}.
\end{equation}

Finally, $\eta$ can be estimated from the equation:
\begin{align}
    \eta
    &\approx0.75\left(\frac{v_\mathrm{out}}{n_\mathrm{e}r_\mathrm{out}}\right)
    \left(\frac{L_\mathrm{H\alpha,b}}{L_\mathrm{H\alpha,b}+L_\mathrm{H\alpha,n}}\right) \\
    &\approx0.75\left(\frac{v_\mathrm{out}}{n_\mathrm{e}r_\mathrm{out}}\right)
    \left(\frac{BNR}{1+BNR}\right),
\label{eq:eta_final}
\end{align}
where $BNR$ are the flux ratio of outflow and narrow line component assuming the same dust-attenuation for the both components.
The values of $\vout$, $\nel$, and $r_\mathrm{out}$ are in the units of km s$^{-1}$, cm$^{-3}$, and kiloparsec, respectively.
For the \Oiii~lines, we adopt Equation \ref{eq:eta_final} in the same manner as the H$\alpha$ lines.
One can derive the equation for the \Oiii emission using the emissivity and abundance of oxygen atom \citepalias[e.g.,][]{Carniani+23} but the results are comparable considering the large uncertainty of $\eta$ estimations. 

The estimation of $\eta$ from emission is largely affected by $\nel$ and $\rout$.
\cite{Isobe+23b} identify an increasing trend of $\nel$ up to $z\sim9$ using NIRSpec data, which may be caused by higher density in the early universe.
We take the median value of $1000~\mathrm{cm^{-3}}$ at $z=8$ and adopt the power-law relation obtained in \cite{Isobe+23b} to estimate $\nel$ for each galaxy.
The effective radius can be approximations of $\rout$ adopted by similar studies including \Car, which is consistent with the rare detection of spatially extended emission associated with broad emission lines \citep[][]{Zhang+23}.
We estimate $\rout$ from the empirical relation between the effective radius and $\Muv$ \citep[][]{Ono+23}. 
Note that our estimates of $\eta$ are smaller than those estimated by \Car~for extremely compact object with small effective radius $\sim100~\mathrm{pc}$.
It is unknown whether the effective radius is similar to $\rout$ in this case due to the projection in the line-of-sight direction.
We therefore adopt the $\rout-\Muv$ relation because we are interested at the mass dependence of $\eta$ instead of the size dependence.

We compare with the $\eta$ estimations of $z\sim1$ studies and the predictions of simulations in Figure \ref{fig:eta_vcir}.
Motivated by the scenario of energy-driven and momentum-driven outflows \citep[e.g.,][]{Muratov+15}, we show the relation between $\eta$ and $\vcir$.
{Our galaxies have $\eta<1$ comparable to star forming galaxies at $z\sim1$ \citep[][]{Swinbank+19}. 
\cite{Ginolfi+20} study $4<z<6$ star forming galaxies and reach a similar conclusion with $\eta$ below or consistent with unity.
\cite{RDP24} observe a $M_*=5.5\times10^9~\mathrm{M_\odot}$ star-forming galaxy at $z\sim3.7$ with JWST NIRSpec IFU. The spatially resolved outflow is find to be inefficient with $\eta\sim0.04$. } 
Our results are largely consistent with those of FIRE-2 simulations at $z\sim0-1$.
Although the offsets between our results and those of simulations at $z\sim6$ may be a result of different definitions, our results suggest inefficient feedback at high redshift and a weak redshift evolution of $\eta$.
{Our estimation of $\eta$ is a good reference to match the properties of high-$z$ galaxies (e.g., mass-metallicity relation; \citealt{Li+23,Nakajima+23}) with model predictions.}

% {TODO: more discussions on feedback efficieny of high-z galaxies. Focus on z>8 results. Implicaitons for z>10 and galaxy overabundance.}

\begin{figure}[htb!]
    \centering
    \includegraphics[width=\linewidth]{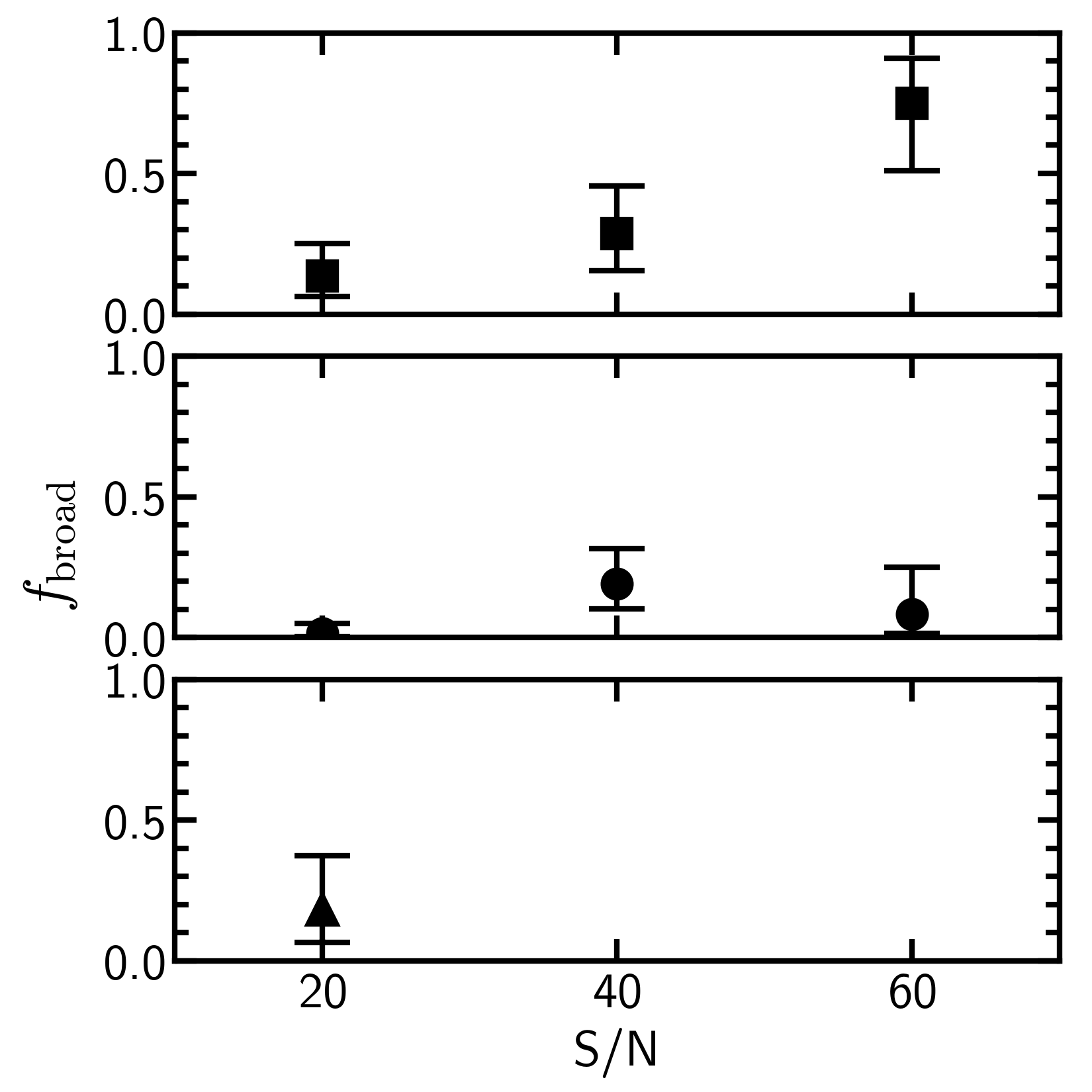}
    \caption{
    Incidence rate of broad outflow emission as a function of redshift measured with the $R\sim2700$ NIRSpec high resolution (Top), $R\sim1000$ medium resolution (Middle), and the $R\sim1600$ NIRCam WFSS (bottom) data. The errors are given by the binomial uncertainties. 
    }
    \label{fig:fbroad_SN}
\end{figure}

\subsection{incidence rate, outflow geometry, and duty-cycle}
\label{rate}

While we identify outflows in 30 out of the 130 galaxies, the detection is subject to the spectral resolution and sensitivity of different observation setup. In Figure \ref{fig:fbroad_SN}, we measure the incidence rate of detecting broad emission lines $f_\mathrm{broad}$ in the S/N bins of $[10,30)$, $[30, 50)$, and $[50, 70)$ for the three typical spectral resolutions. The incidence rates obtained by NIRSpec medium resolution are significantly lower those obtained by high resolution, likely due to the limited spectral resolution. 
\noedit{Actually, if we convolve the high resolution spectra with LFS to have consistent spectral resolution as the medium resolution spectra, we obtain a detection rate of $\sim10\%$ consistent with the one of medium resolution spectra.}
Such a selection bias is consistent with the fact that $\vout$ measured with NIRSpec medium resolution tends to be larger than those measured with high resolution. In the S/N bin of $[10,30)$, data taken by NIRSpec high resolution and NIRCam WFSS has comparable incidence rates of $\sim14\%$ and $\sim18\%$, respectively, which may imply $R>1600$ is sufficient for the detection of outflows. In the case of NIRSpec high resolution, the incidence rate increase from $\sim14\%$ to $\sim75\%$ with higher S/N. Note that the increasing S/N degenerates with the apparent luminosity of the galaxy which relates to the SFR and stellar mass. An overall detection rate of $\sim30\%=20/65$ using the NIRSpec high resolution data would be the fiducial value with the current data, which provides some hints for the physical picture of outflows.

\begin{figure}[htb!]
    \centering
    \includegraphics[width=\linewidth]{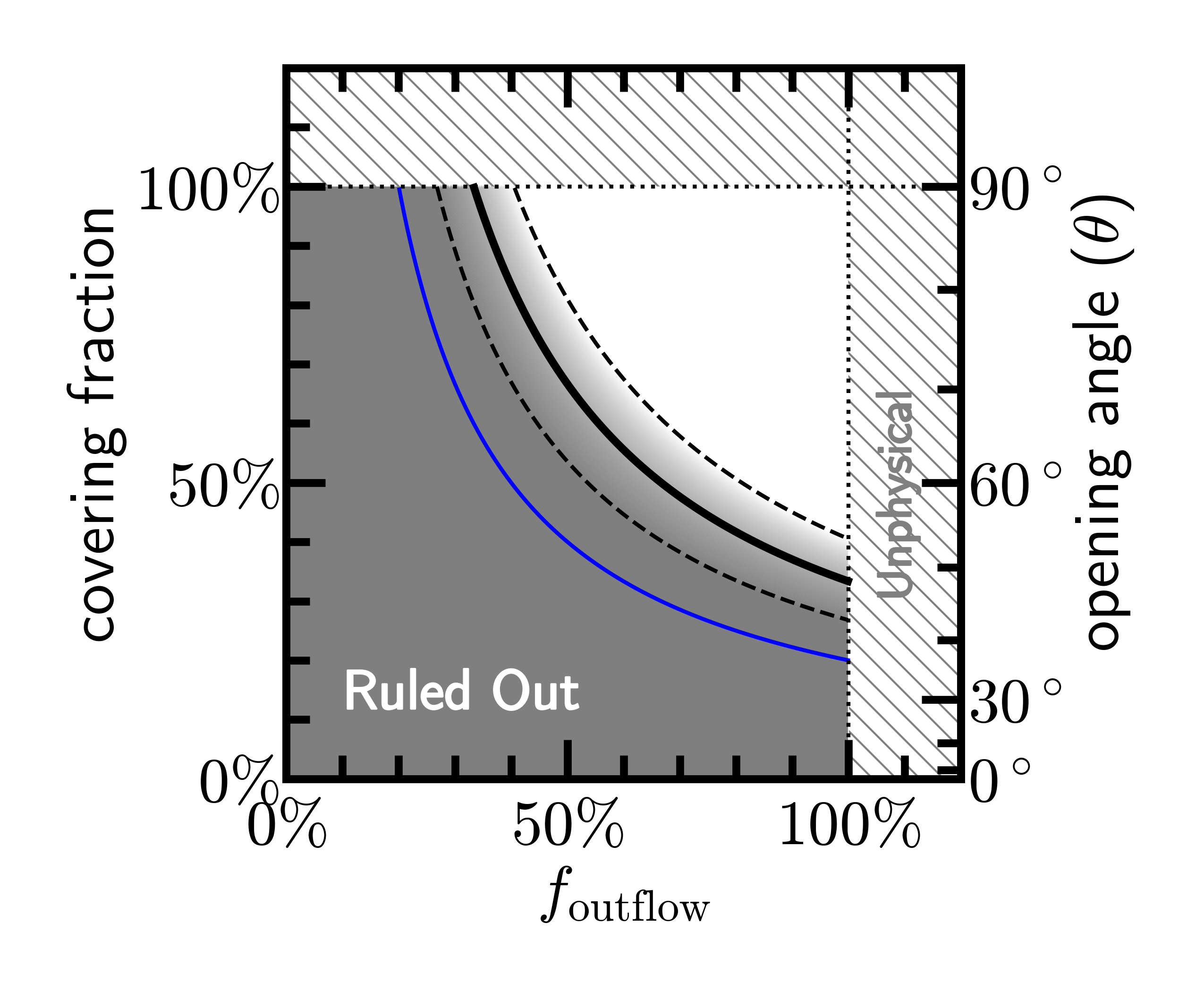}
    \caption{
    Incidence rate of broad outflow emission given by the product of covering fraction and outflow duty-cyle. The hatched region are unphysical while the grey shaded region are ruled out by an incidence rate of 30\% shown with the black solid line. The black dashed line and the shaded gradient in-between are given by the binomial uncertainties. The blue dotted line corresponds to the incidence rate in local galaxies estimated from the results of \cite{Stuber+21}.
    }
    \label{fig:outflow_rate}
\end{figure}

Two important aspects of outflows that remain unclear are the geometry and the duty-cycle.
Interestingly, these two factors simultaneously determine the incidence rate of outflow presented as broad emission lines: 
\begin{equation}
f_\mathrm{broad}=f_\mathrm{outflow}\times\frac{\Omega}{4\pi},
\end{equation} where $f_\mathrm{outflow}$ is the fraction of outflowing phase during the current star formation (hereafter outflow rate). $\Omega$ and $\theta$ are the solid angle and half-opening angle of the outflow cone, respectively.
We consider the outflow cone enclosing the directions where the outflowing gas is visible as broad component in line-of-sight direction, which is somewhat different from the common definition using outflowing medium \cite[e.g.,][]{Obied+16}.
In Figure \ref{fig:outflow_rate}, we show how outflow rate and opening angle degenerate with each other. The observed incidence rate rules out small opening angle and $f_\mathrm{outflow}$.
We obtain a detection rate of $\sim30\%=20/65$ using the NIRSpec high resolution data, suggestive of a wide half-opening angle of $\gtrsim45$ deg and a frequent duty-cycle of $f_\mathrm{outflow}\gtrsim 30\%$.
This is consistent with the distribution of outflow galaxies in Figure \ref{fig:SFR_Mstar} that broadly agrees with the distribution of parent sample and the $z\sim6$ MS, suggestive of the ubiquity of outflows.
\cite{Stuber+21} report a detection rate of $\sim25\%$ for outflows in nearby main-sequence galaxies selected with relatively low inclination ($i<75^\circ$).
After correcting for the inclination selection, the fraction of line-of-sight outflow is $\sim20\%=25\%/(1 - \cos75^\circ)$ in nearby main-sequence galaxies which is smaller than the one we obtain, suggestive of higher frequency or wider opening angle for high-$z$ outflows.
Assuming a constant opening angle of $45^\circ$ \citep[e.g.,][]{Venturi+18} for both nearby and high-$z$ galaxies, the outflow rate is $\sim60\%$ for nearby galaxies but as high as 100\% for high-$z$ galaxies as shown in Figure \ref{fig:outflow_rate}.
We note our constraints based on the detection rate of $\sim30\%$ are lower limits given by the sensitivity and resolution of NIRSpec data.
\noedit{We also recognize that secondary emission line components can also be caused by physical processes other than outflows such as galaxy-galaxy interaction \citep[e.g.,][]{Maschmann+19}, expanding ionized bubbles, or gas inflows \citep[e.g.,][]{Lena+15}. Although the majority of our galaxies have compact and isolated morphologies, different sources of turbulent motions cannot be ruled out with the current dataset.
Based on the assumption that broad emission lines are mostly contributed by outflows, we summarize that} outflows in the early universe are likely to be more spherical and frequent, which is to be tested with future observations.

Another interesting interpretation can be made considering the galaxies with AGN signatures. The detection rate of outflows for Type-1 and Type-2 AGNs are $4/12=33\%$ and $1/7=15\%$, respectively. Note that Type-2 AGNs are all observed with high resolution spectra while the detection rate of outflows are smaller than those of Type-1 AGNs and full sample of high resolution spectra. Because Type-2 AGNs are considered to be edge on in the standard AGN unification scheme \citep[][]{Urry+95}, the outflows are likely stronger in the polar direction and harder to be identified from line-of-sight velocities.

% \section{Implications on Early Galaxy Formation}

\section{Summary}
\label{summary}

We study feedback of star formation and AGN activity in 130 galaxies with $-22<M_{\rm UV}<-16$ at $z=3-9$ identified in JWST NIRSpec and NIRCam WFSS data taken by the ERO, CEERS, FRESCO, GLASS, and JADES programs. We detect outflows powered by inefficient feedback that are frequent and fountain-like with wide opening angles. We summarize our results as below:
\begin{enumerate}
    \item We identify 30 out of the 130 galaxies via broad components of FWHM$\sim200-700~\mathrm{km~s^{-1}}$ in the emission lines of H$\alpha$ and \Oiii~that trace ionized outflows. Four out of the 30 outflowing galaxies are Type-1 AGNs while one galaxy hosts a Type-2 AGN. 
    \item We obtain $\sim 80-500~\mathrm{km~s^{-1}}$ for the outflow velocities. \noedit{We find that the outflow velocities are slower than low-z galaxies with similar SFRs, which may be explained by the low stellar masses of high-z galaxies.} The outflow velocities of AGN are large but not significantly different from the others.
    \item The outflow velocities are typically not high enough to escape from the galactic potentials, possibly suggesting fountain-type outflows, which are concluded on the basis of thorough comparisons with recent JWST results.
    \item \noedit{Galaxies with SFRs larger than MS host faster and more prominent outflows, which agrees with the picture that outflows suppress the star formation in galaxies during a bursty phase.} We estimate mass loading factors $\eta$ to be $\eta=0.1-1$ that are not particularly large, but comparable with those of $z\sim 1$ outflows. 
    \item The large fraction of galaxies with outflows (30\% with high resolution data) provides constraints on outflow parameters, suggesting a wide opening angle of $\gtrsim 45$ deg and a large duty-cycle of $\gtrsim30$\%, which gives a picture of more frequent and spherical outflows in high-$z$ galaxies.
\end{enumerate}  

We thank S. Carniani for sharing the details of data analysis and the useful discussions.
We thank D. Kashino for helping with the data reductions.
We thank K. Nagamine, A.K. Inoue, and T. Hashimoto for having useful discussions.
We thank F. Sun for discussions on objects selected from FRESCO.
We are grateful to staff of the James Webb Space Telescope Help Desk for letting us know useful information.
This work is based on observations made with the NASA/ESA/CSA James Webb Space Telescope.
The data were obtained from the Mikulski Archive for Space Telescopes at the Space Telescope Science Institute, which is operated by the Association of Universities for Research in Astronomy, Inc., under NASA contract NAS 5-03127 for JWST.
These observations are associated with programs 1180, 1895, 2736, 1324, and 1345.
The authors acknowledge the ERO, GLASS, and CEERS teams led by Klaus M. Pontoppidan, Tommaso Treu, and Steven L. Finkelstein, respectively, for developing their observing programs with a zero-exclusive-access period.
This work is based on observations taken by the CANDELS Multi-Cycle Treasury Program with the NASA/ESA HST, which is operated by the Association of Universities for Research in Astronomy, Inc., under NASA contract NAS5-26555.
This work was supported by the joint research program of the Institute for Cosmic Ray Research (ICRR), University of Tokyo. 
This publication is based upon work supported by the World Premier International Research Center Initiative (WPI Initiative), MEXT, Japan.
M.O., K.N., Y.H., and Y.I. are supported by JSPS KAKENHI Grant Nos. 20H00180/21H04467, 20K22373, 21K13953, and 21J20785,  respectively.
\software{astropy \citep{astropy:2013,astropy:2018}, SExtractor \citep{sextractor}, galpy \citep{galpy}}

\bibliography{ref}{}
\bibliographystyle{aasjournal}

\begin{deluxetable*}{ccCCCCcCCC}
    % \tablenum{2}
    \tablecaption{Galaxy and Outflow Properties}
    \tablewidth{0pt}
    \tablehead{
        \colhead{ID} & \colhead{Redshift} &
        \colhead{$\Muv$} & \colhead{$\log(M_*/\mathrm{M_\odot})$} &
        \colhead{$\log(\mathrm{SFR/M_\odot~yr^{-1}})$} & \colhead{$\vcir$} & 
        \colhead{line} & \colhead{$\vout$} & \colhead{$\eta$} & \colhead{$\Delta\mathrm{AIC}$}\\
        \nocolhead{} & \nocolhead{} & 
        \nocolhead{} & \nocolhead{} &
        \nocolhead{} & \colhead{($\mathrm{km~s^{-1}}$)} &
        \nocolhead{} & \colhead{($\mathrm{km~s^{-1}}$)} & \nocolhead{} & \nocolhead{} \\
        \colhead{(1)} & \colhead{(2)} & 
        \colhead{(3)} & \colhead{(4)} & 
        \colhead{(5)} & \colhead{(6)} & 
        \colhead{(7)} & \colhead{(8)} & \colhead{(9)} & \colhead{(10)}
    }
    % \decimalcolnumbers
    \startdata
    CEERS 00397 & 6.00 & -21.23^{+0.17}_{-0.24} & 8.45^{+0.53}_{-0.06} & 1.76^{+0.02}_{-0.03} & 271.23^{+17.16}_{-24.23} & [{\sc Oiii}] & 222\pm94 & -0.85 & -4.32 \\
CEERS 00698 & 7.47 & -21.60\pm0.11 & 9.39\pm0.17 & 1.71^{+0.14}_{-0.21} & 313.63\pm13.76 & [{\sc Oiii}] & 441\pm42 & -0.29 & -74.46 \\
CEERS 01149 & 8.18 & -20.83\pm0.19 & 8.65\pm0.33 & 1.40^{+0.15}_{-0.22} & 245.48\pm15.76 & [{\sc Oiii}] & 226\pm56 & -0.50 & -14.53 \\
CEERS 01244 & 4.48 & -19.50\pm0.18 & 8.62\pm0.88 & 2.36^{+0.15}_{-0.22} & 164.58\pm7.02 & H$\alpha$ & 639\pm143 & -0.12 & -440.32 \\
CEERS 01561 & 6.20 & -19.91\pm0.13 & 8.70\pm0.83 & 1.28^{+0.15}_{-0.22} & 182.11\pm5.90 & H$\alpha$ & 245\pm65 & -0.49 & -3.36 \\
  &   &   &   &   &   & [{\sc Oiii}] & 367\pm122 & -0.57 & -3.73 \\
ERO 04590 & 8.50 & -17.95\pm0.23 & 7.60^{+0.63}_{-0.51} & 0.66\pm0.03 & 133.80\pm4.17 & [{\sc Oiii}] & 378\pm79 & -0.50 & -7.60 \\
ERO 06355 & 7.67 & -20.33\pm0.00 & 8.77^{+0.08}_{-0.01} & 1.41\pm0.01 & 203.87\pm0.00 & [{\sc Oiii}] & 386\pm33 & -0.59 & -115.87 \\
ERO 10612 & 7.66 & -20.07\pm0.08 & 7.78^{+0.29}_{-0.03} & 1.14^{+0.01}_{-0.02} & 189.65\pm4.01 & [{\sc Oiii}] & 254\pm35 & -0.70 & -47.50 \\
FRESCO N 1 & 7.50 & -21.37\pm0.04 & 9.82\pm0.49 & 1.41\pm0.31 & 286.82\pm4.27 & [{\sc Oiii}] & 476\pm68 & -0.39 & -19.59 \\
FRESCO S 3 & 8.02 & -20.10\pm0.15 & 9.04\pm0.50 & 0.69\pm0.47 & 195.01\pm7.95 & [{\sc Oiii}] & 366\pm142 & -0.10 & -5.61 \\
GLASS 10000 & 7.88 & -20.36^{+0.16}_{-0.14} & 8.16^{+0.27}_{-0.07} & 1.25^{+0.03}_{-0.04} & 208.20^{+9.75}_{-8.53} & [{\sc Oiii}] & 143\pm8 & -0.31 & -59.34 \\
GLASS 100001 & 7.87 & -20.29^{+0.50}_{-0.29} & 9.15^{+0.55}_{--0.00} & 0.95^{+0.10}_{-0.13} & 203.94^{+29.14}_{-16.90} & [{\sc Oiii}] & 157\pm42 & -0.45 & -5.96 \\
GLASS 100003 & 7.88 & -20.69^{+0.10}_{-0.17} & 8.27^{+0.38}_{-0.04} & 1.22^{+0.04}_{-0.05} & 230.48^{+7.48}_{-12.72} & [{\sc Oiii}] & 317\pm69 & -0.53 & -12.18 \\
GLASS 110000 & 5.76 & -19.91^{+0.64}_{-0.34} & 9.22^{+0.16}_{-0.27} & 0.53^{+0.04}_{-0.05} & 176.48^{+28.14}_{-14.95} & [{\sc Oiii}] & 139\pm4 & -0.58 & -120.65 \\
GLASS 150029 & 4.58 & -19.20^{+0.10}_{-1.08} & 9.12^{+0.03}_{-0.33} & 1.04\pm0.02 & 141.08^{+3.14}_{-33.97} & H$\alpha$ & 114\pm16 & -1.17 & -784.13 \\
  &   &   &   &   &   & [{\sc Oiii}] & 83\pm8 & -1.07 & -92.33 \\
GLASS 160133 & 4.01 & -18.94^{+0.08}_{-0.09} & 8.11^{+0.41}_{-0.22} & 1.16\pm0.01 & 139.22^{+2.31}_{-2.60} & H$\alpha$ & 183\pm9 & -0.84 & -2006.51 \\
  &   &   &   &   &   & [{\sc Oiii}] & 227\pm6 & -1.05 & -843.81 \\
GLASS 40066 & 4.02 & -20.43^{+1.86}_{-0.26} & 9.40^{+0.33}_{-0.10} & 1.57\pm0.02 & 205.73^{+134.98}_{-18.87} & [{\sc Oiii}] & 222\pm27 & -1.22 & -21.39 \\
GLASS 50038 & 5.77 & -20.12^{+1.36}_{-0.19} & 9.09^{+0.39}_{-0.09} & 0.91^{+0.04}_{-0.05} & 186.47^{+68.25}_{-9.54} & [{\sc Oiii}] & 176\pm5 & -0.70 & -253.11 \\
GLASS 80029 & 3.95 & -18.92^{+1.00}_{-0.50} & 8.50^{+0.22}_{-0.49} & 0.47^{+0.04}_{-0.05} & 137.77^{+28.51}_{-14.26} & [{\sc Oiii}] & 113\pm14 & -1.10 & -34.72 \\
GLASS 80070 & 4.80 & -17.77^{+1.01}_{-0.19} & 7.42^{+0.70}_{-0.04} & 0.17^{+0.04}_{-0.05} & 109.00^{+19.09}_{-3.59} & [{\sc Oiii}] & 134\pm4 & -0.82 & -168.30 \\
JADES 00004297 & 6.71 & -18.49\pm0.07 & 8.24\pm0.49 & -0.11\pm0.03 & 130.29\pm1.45 & [{\sc Oiii}] & 177\pm46 & -0.98 & -2.19 \\
JADES 00005457 & 4.86 & -18.06\pm0.18 & 7.69\pm0.05 & 0.09\pm0.19 & 115.29\pm3.80 & [{\sc Oiii}] & 333\pm72 & -0.56 & -6.13 \\
JADES 00007892 & 4.23 & -18.74\pm0.04 & 7.81\pm0.04 & 0.09\pm0.01 & 136.54\pm0.97 & H$\alpha$ & 127\pm25 & -0.86 & -16.35 \\
JADES 00007938 & 4.81 & -19.53\pm0.06 & 8.73\pm0.49 & 0.57\pm0.23 & 155.50\pm2.14 & H$\alpha$ & 173\pm38 & -1.03 & -5.59 \\
  &   &   &   &   &   & [{\sc Oiii}] & 133\pm48 & -1.00 & -3.78 \\
JADES 00008083 & 4.65 & -18.75\pm0.06 & 8.45\pm0.03 & 0.19\pm0.23 & 129.12\pm1.60 & H$\alpha$ & 170\pm28 & -0.96 & -366.66 \\
JADES 00018090 & 4.78 & -19.17\pm0.05 & 7.85\pm0.05 & 0.61\pm0.02 & 142.39\pm1.67 & H$\alpha$ & 102\pm13 & -0.69 & -9.49 \\
JADES 00020961 & 7.04 & -18.90\pm0.05 & 8.41\pm0.49 & 0.10\pm0.26 & 142.25\pm1.27 & [{\sc Oiii}] & 141\pm30 & -0.73 & -5.03 \\
JADES 10009506 & 3.60 & -19.33\pm0.03 & 8.71\pm0.49 & 0.53\pm0.23 & 145.00\pm0.83 & H$\alpha$ & 97\pm16 & -0.90 & -8.68 \\
JADES 10013704 & 5.92 & -19.02\pm0.07 & 8.63\pm0.02 & 1.14\pm0.02 & 147.93\pm1.95 & [{\sc Oiii}] & 180\pm51 & -0.96 & -4.39 \\
JADES 10035295 & 3.59 & -18.59\pm0.08 & 8.35\pm0.49 & 0.16\pm0.23 & 124.15\pm1.85 & H$\alpha$ & 93\pm22 & -1.05 & -4.59

    \enddata
    \tablecomments{Columns: (1) ID. (2) Redshift. (3)-(5) UV magnitude, stellar mass, and star formation rate of the galaxy. (6) circular velocity of the DM halo hosting the galaxy (7) emission line used to trace outflows (8) outflow velocity estimated with Equation \ref{eq:vout}. (9) mass-loading factor estimated with Equation \ref{eq:eta}. (10) difference of AIC between the two-component model and one-component model (three-component model and two-component model for the case of Type-1 AGNs)}
    \label{tab:outflow}
\end{deluxetable*}

% \iffalse
\newpage
\appendix
\restartappendixnumbering

\section{Galaxies identified in FRESCO}

\begin{figure*}[tbh!]
    \centering
    \includegraphics[height=0.25\textheight]{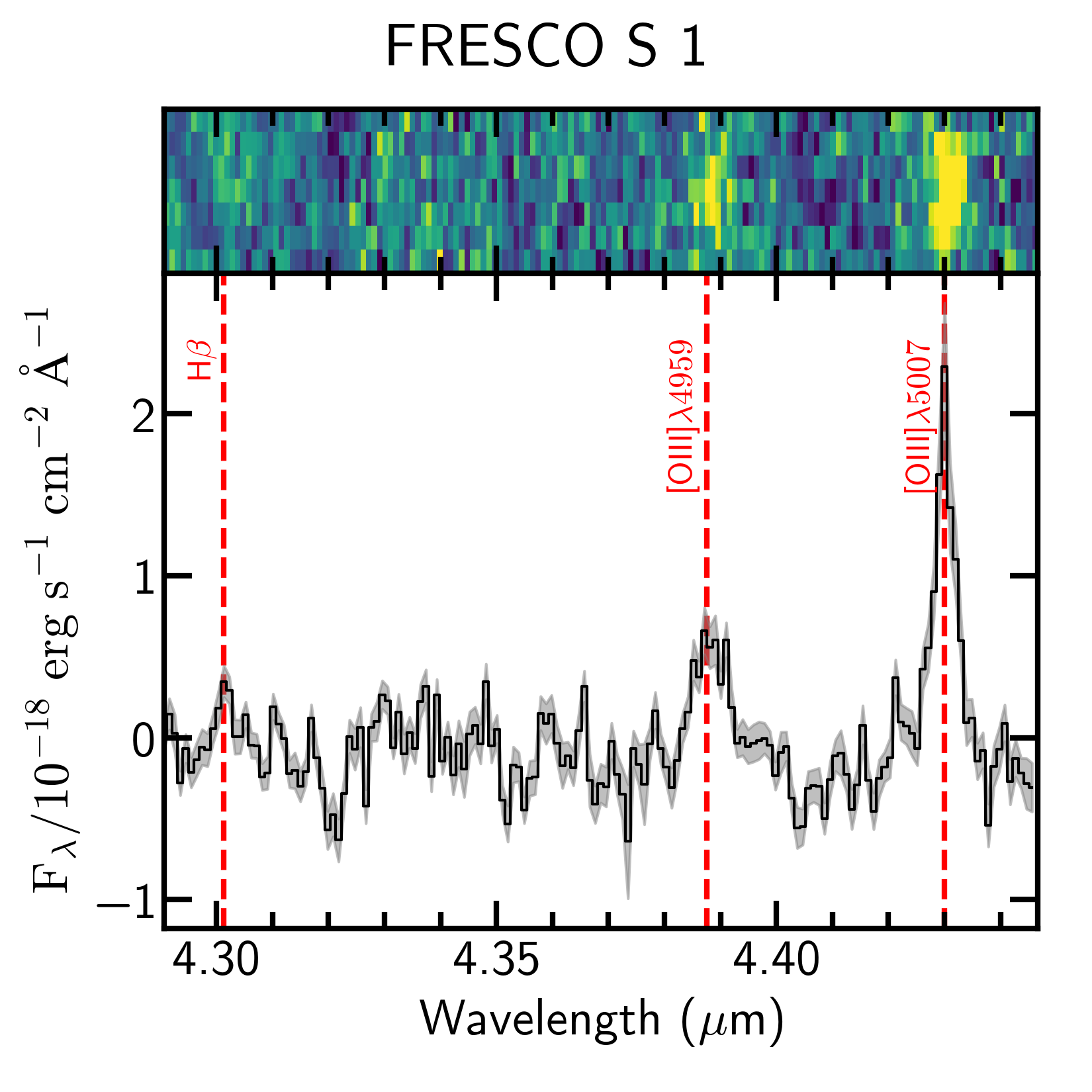}%
    \includegraphics[height=0.25\textheight]{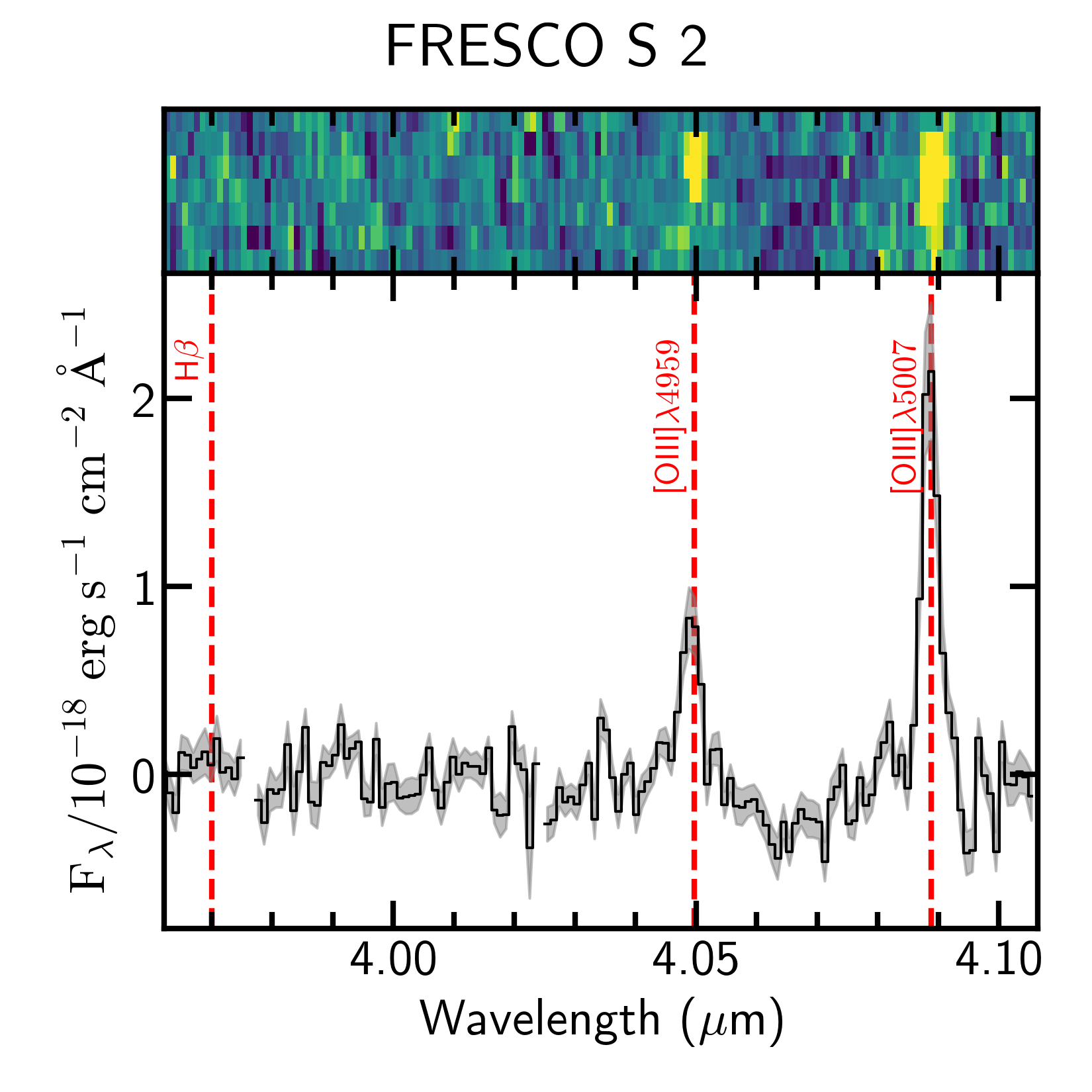}\\
    \includegraphics[height=0.25\textheight]{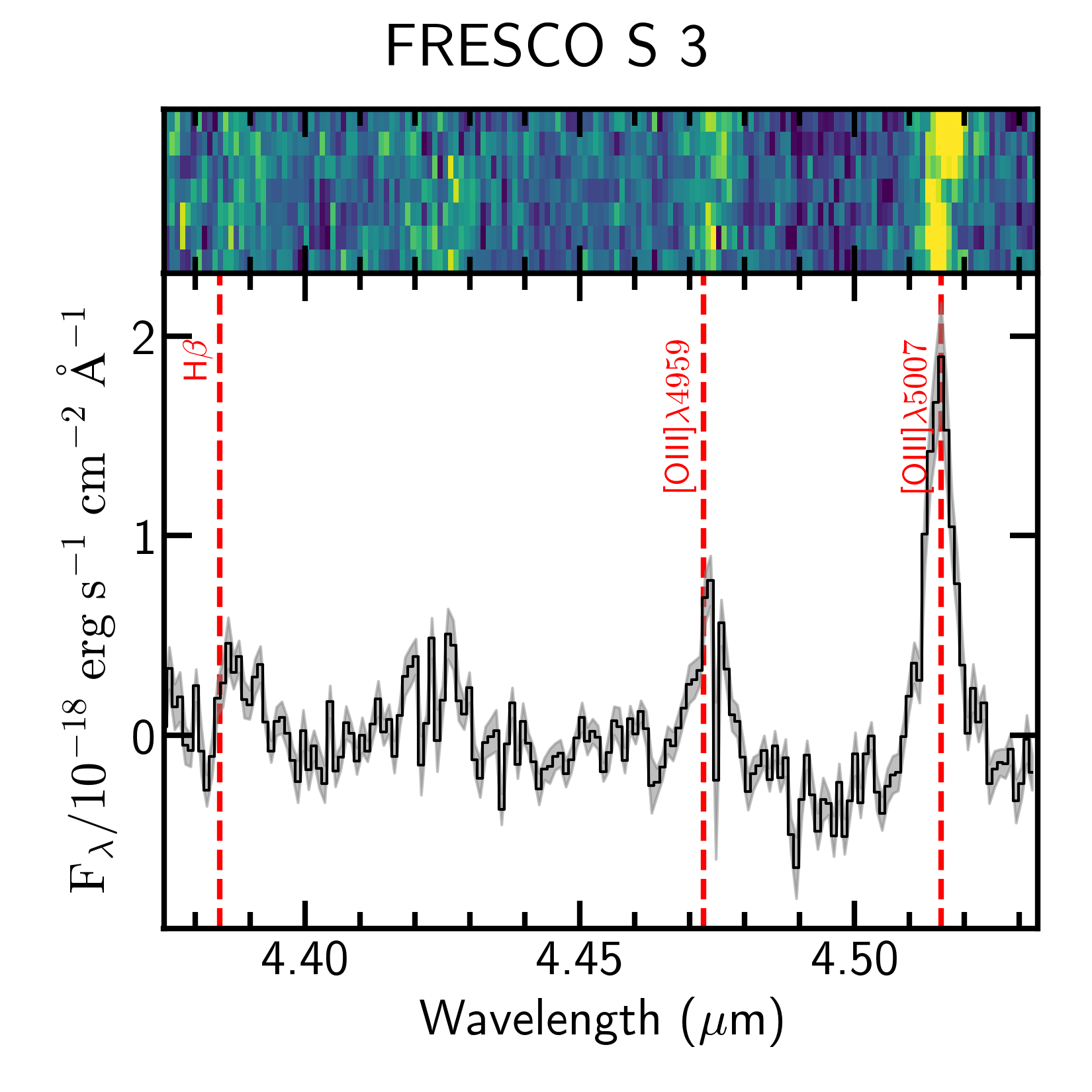}%
    \includegraphics[height=0.25\textheight]{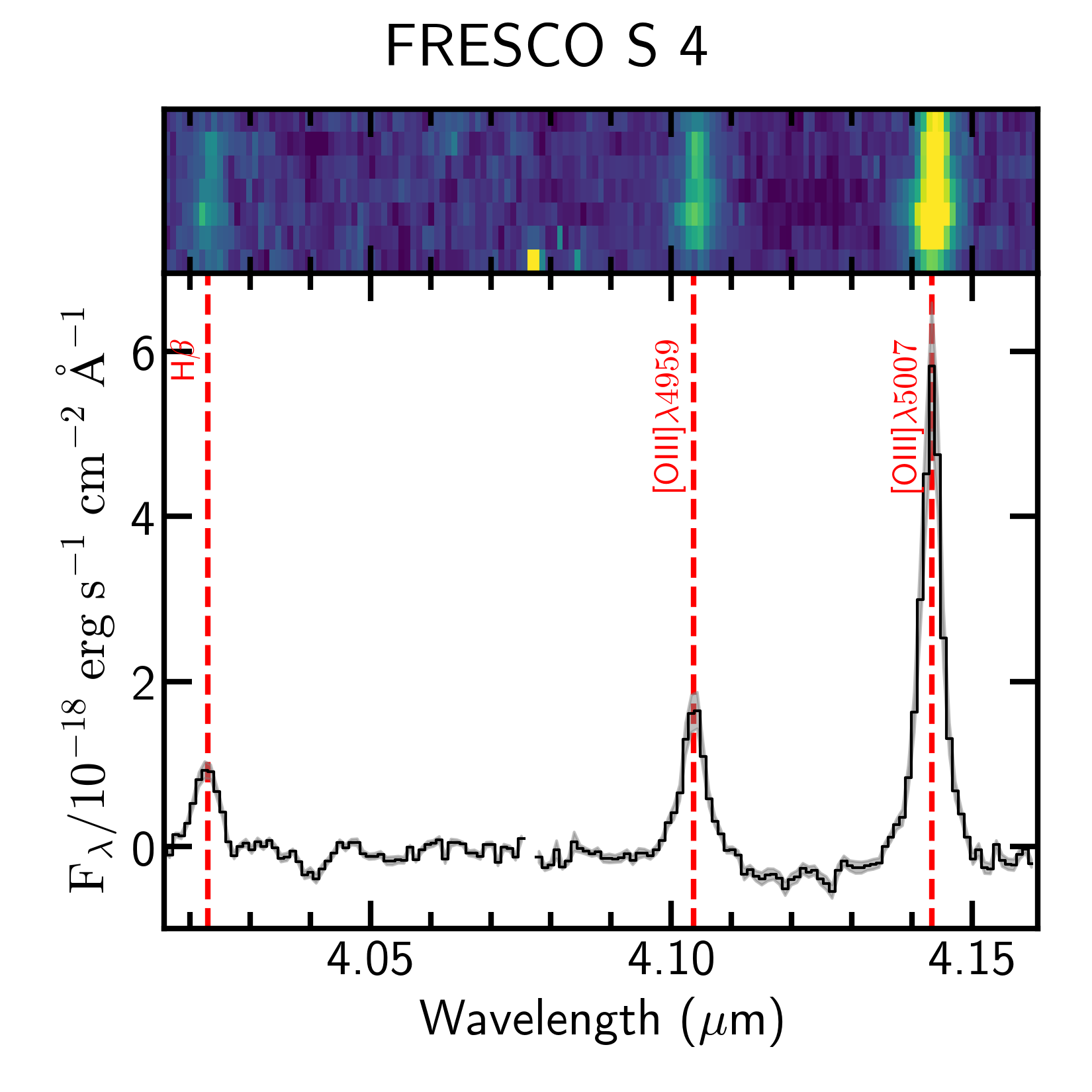}\\
    \includegraphics[height=0.25\textheight]{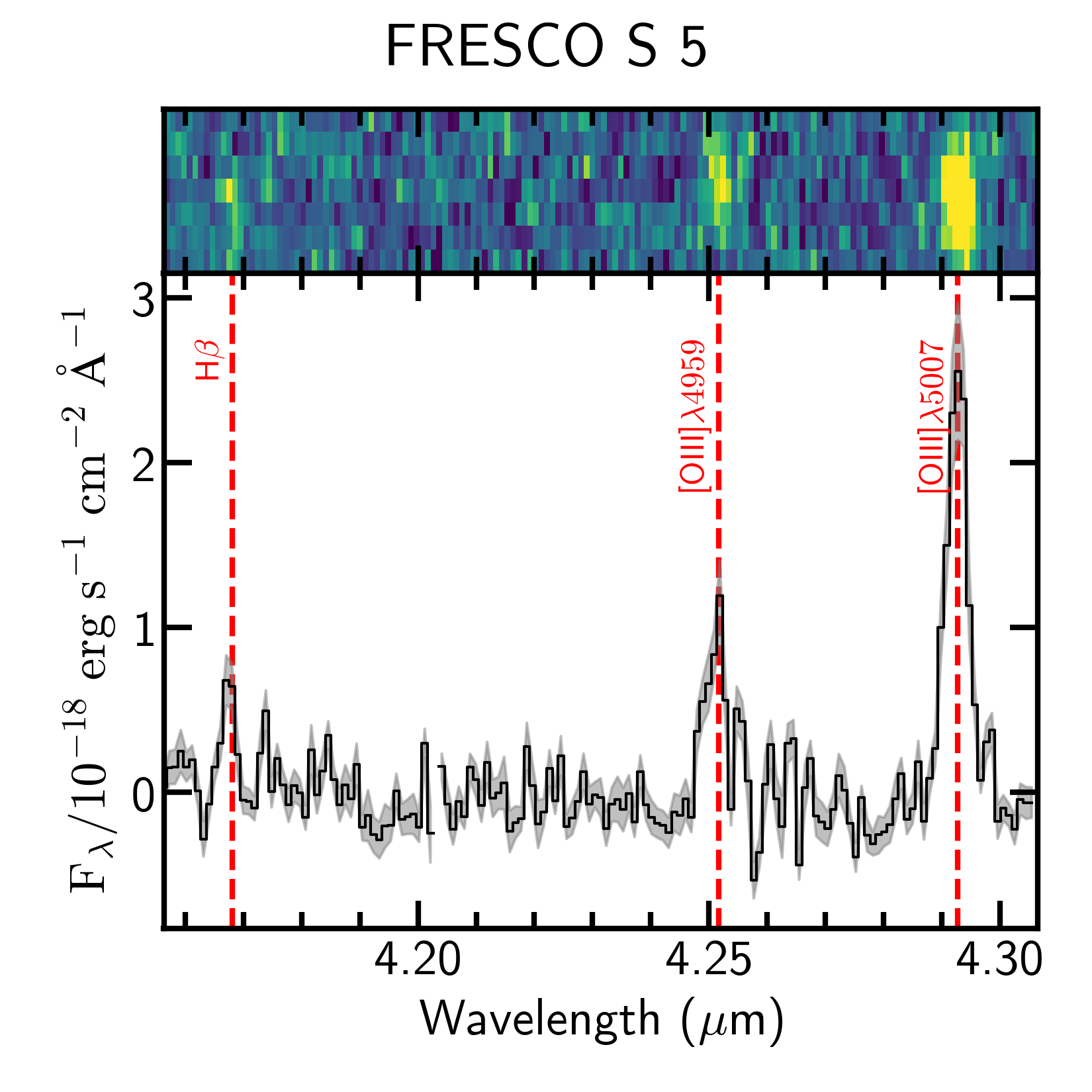}%
    \includegraphics[height=0.25\textheight]{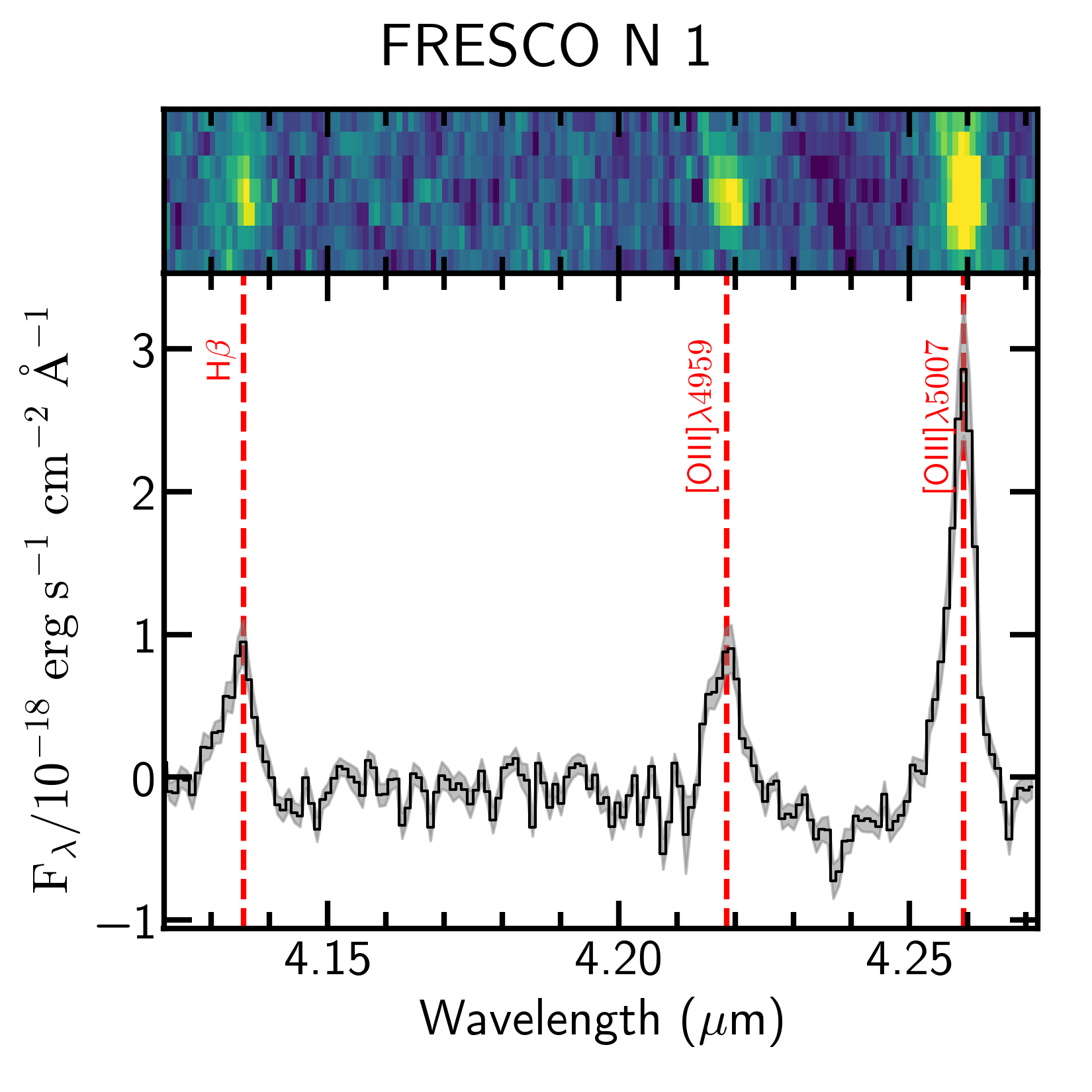}\\
    \caption{
    Spectra of \Oiii~emitters found in the FRESCO data. The top panels show the 2d spectrum extracted with an aperture radius of 3 pixels. The bottom panels show the 1d spectrum and flux noise.
    }
    \label{fig:FRESCO_2dspec}
\end{figure*}

\begin{figure*}[thb!]
\centering
\includegraphics[height=0.25\textheight]{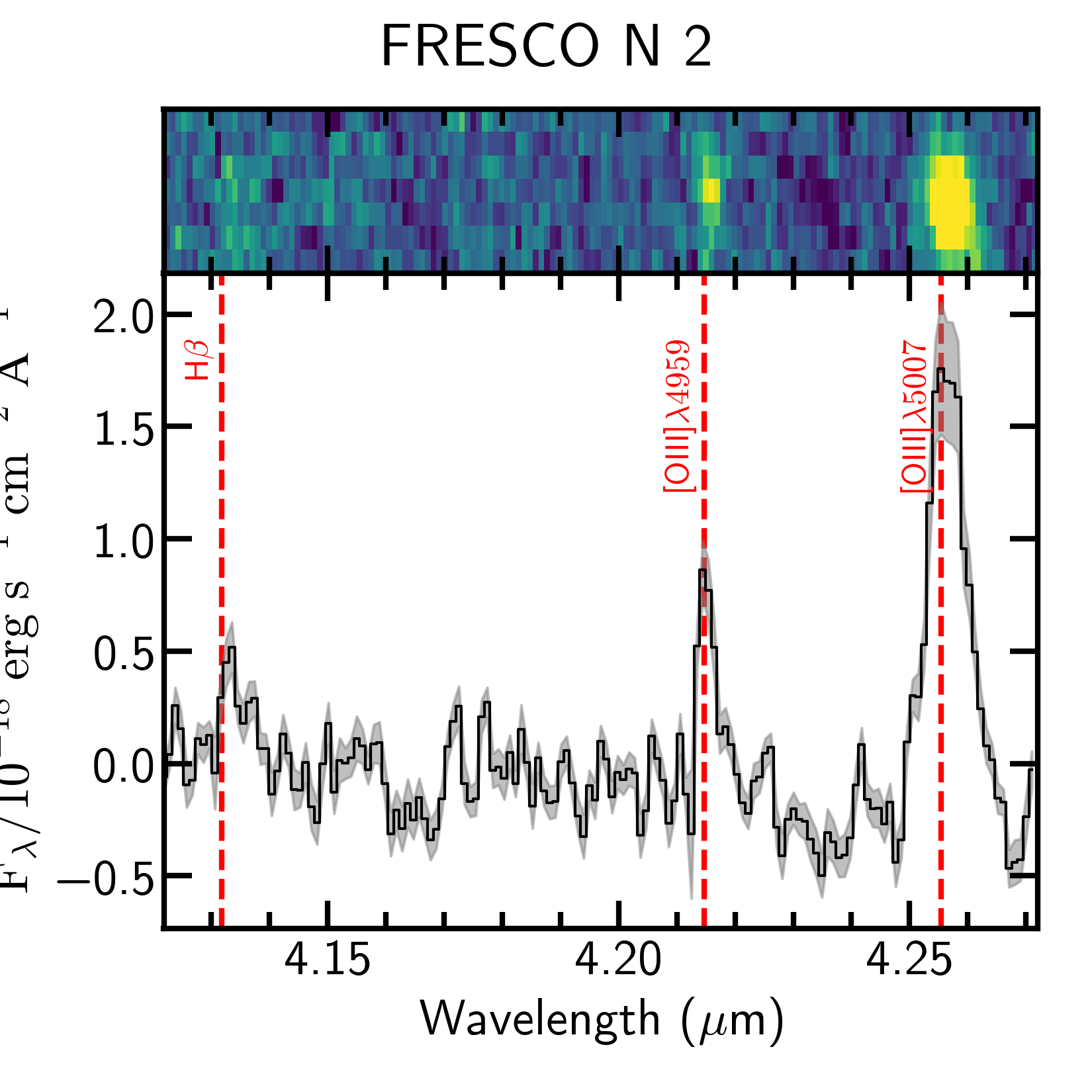}%
\includegraphics[height=0.25\textheight]{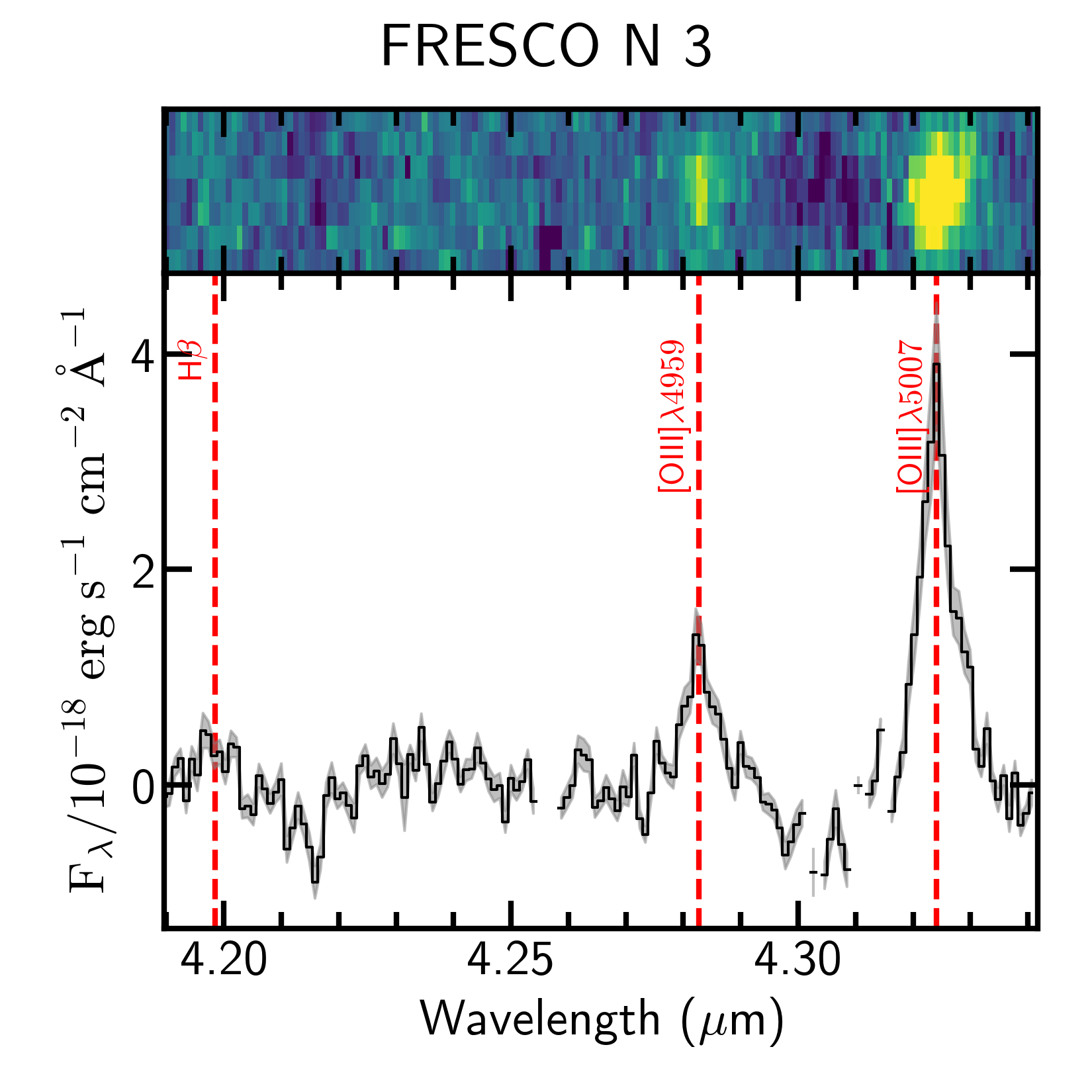}\\
\includegraphics[height=0.25\textheight]{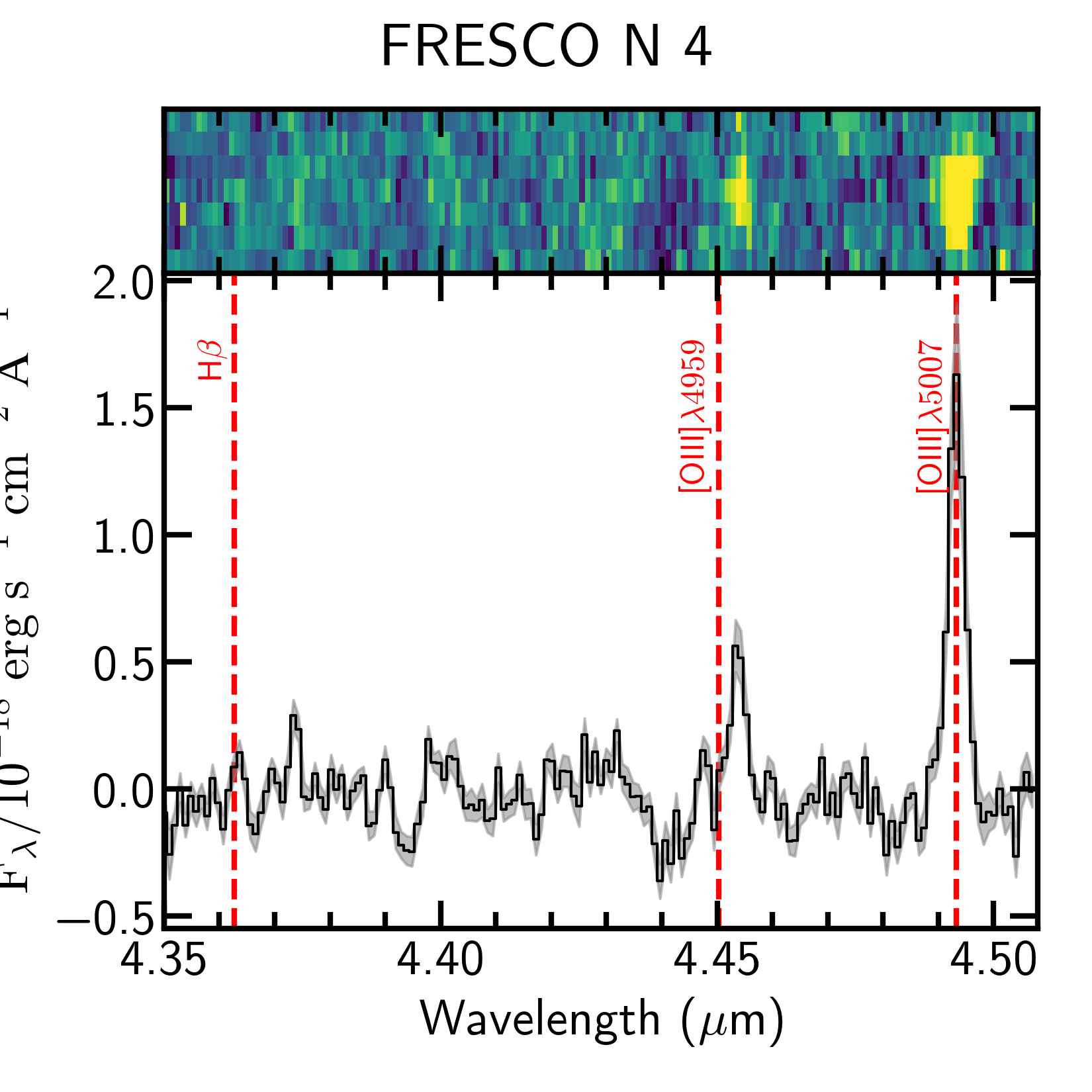}%
\includegraphics[height=0.25\textheight]{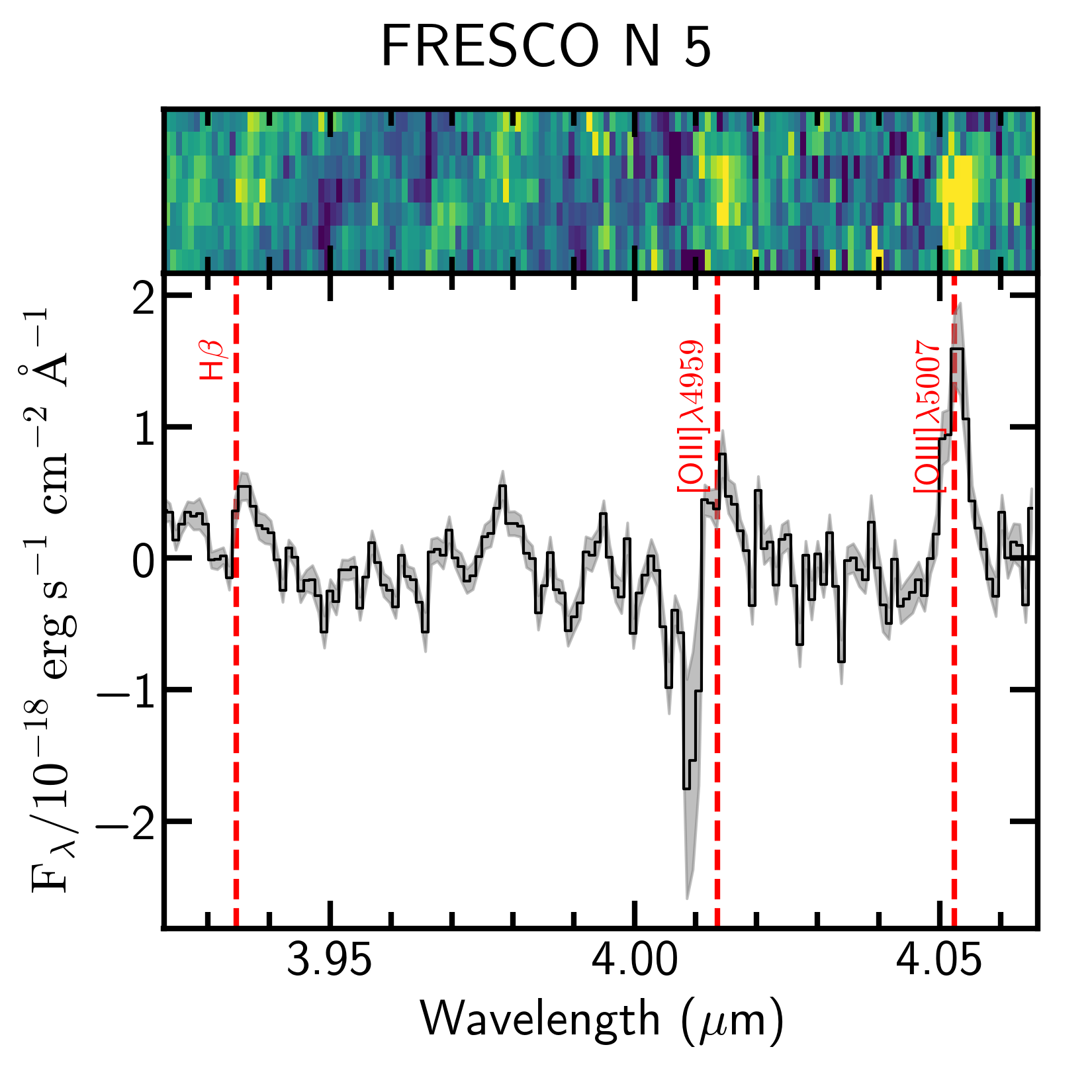}\\
\includegraphics[height=0.25\textheight]{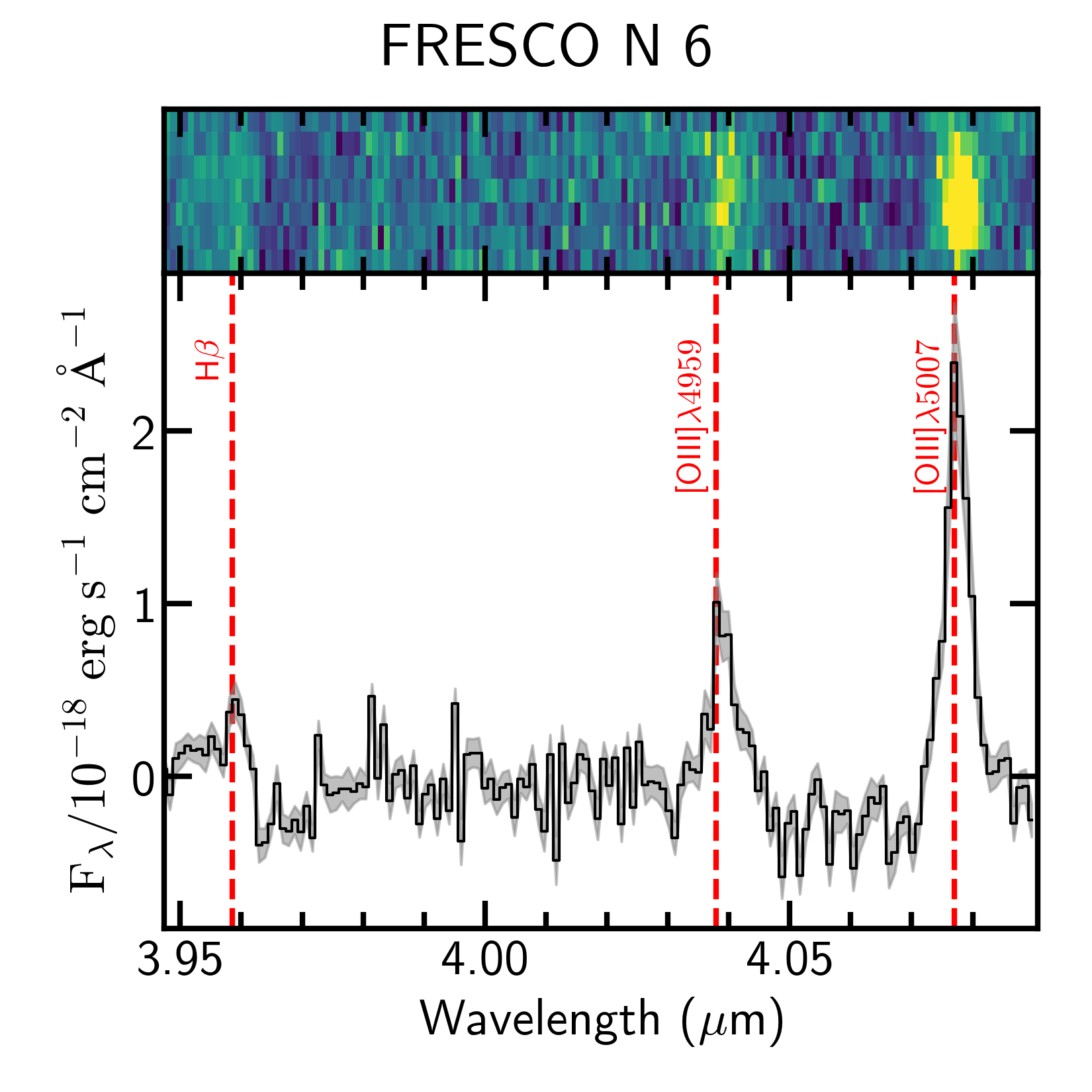}%
\includegraphics[height=0.25\textheight]{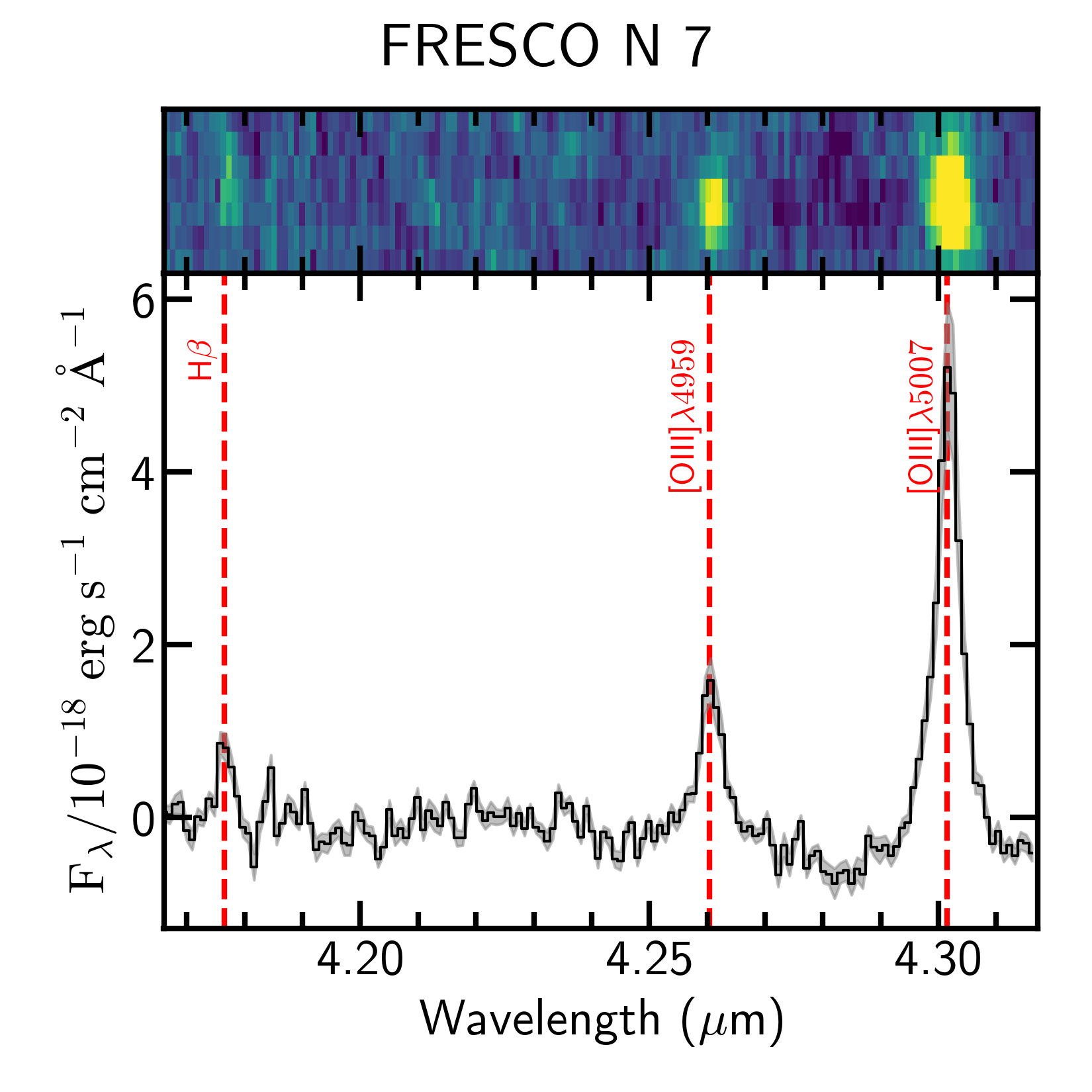}
\caption{
Figure \ref{fig:FRESCO_2dspec} continued.
}
\label{fig:FRESCO_2dspec_2}
\end{figure*}

\clearpage
\begin{deluxetable*}{ccccc}
    % \tablenum{2}
    \tablecaption{Galaxy sample of FRESCO}
    \tablewidth{0pt}
    \tablehead{
        \colhead{ID} & 
        \colhead{R.A.} & \colhead{Decl.} & \colhead{Redshift} &
        \colhead{$F_{\mathrm{\textsc{[Oiii]}}\lambda5007}$} \\
        \nocolhead{} &
        \nocolhead{} & \nocolhead{} & \nocolhead{} & 
        \colhead{($10^{-18}~\mathrm{erg~s^{-1}~cm^{-2}}$)} \\
        \colhead{(1)} & \colhead{(2)} & 
        \colhead{(3)} & \colhead{(4)} & 
        \colhead{(5)}
    }
    % \decimalcolnumbers
    \startdata
    FRESCO N \#1 & 12:36:37.90 & 62:18:08.78 & 7.505 & $156.52\pm11.56$ \\
    FRESCO N \#2 & 12:36:37.86 & 62:18:08.60 & 7.501 & $162.11\pm10.94$ \\
    FRESCO N \#3 & 12:37:04.80 & 62:17:18.82 & 7.637 & $280.39\pm17.86$ \\
    FRESCO N \#4 & 12:36:54.03 & 62:17:10.67 & 7.974 & $61.85\pm7.40$ \\
    FRESCO N \#5 & 12:36:48.88 & 62:16:06.50 & 7.094 & $71.89\pm9.91$ \\
    FRESCO N \#6 & 12:36:59.91 & 62:14:28.41 & 7.144 & $122.34\pm10.87$ \\
    FRESCO N \#7 & 12:37:19.94 & 62:15:26.04 & 7.591 & $287.28\pm19.54$ \\
    FRESCO S \#1 & 03:32:21.00 & -27:48:53.55 & 7.848 & $104.90\pm10.74$ \\
    FRESCO S \#2 & 03:32:27.25 & -27:47:36.54 & 7.166 & $84.73\pm9.85$ \\
    FRESCO S \#3$^\dag$ & 03:32:32.91 & -27:45:37.45 & 8.019 & $133.22\pm10.10$ \\
    FRESCO S \#4 & 03:32:46.90 & -27:50:07.60 & 7.275 & $248.55\pm17.12$ \\
    FRESCO S \#5 & 03:32:41.42 & -27:44:38.18 & 7.573 & $125.26\pm11.10$
    \enddata
    \tablecomments{The name, coordinates, redshift, and emission line flux of the galaxies identified in the NIRCam WFSS data set of FRESCO. {$^\dag$ Possible $z=1.7$ interloper (F. Sun, private communication)}}
    \label{tab:FRESCO_objects}
\end{deluxetable*}
% \fi

\end{document}